\documentclass[reprint,prd,nofootinbib,superscriptaddress]{revtex4}
\usepackage{amsfonts}
\usepackage{amsmath}
\usepackage{amssymb}
\usepackage[english]{babel}
\usepackage{graphicx}
\usepackage{epsfig}
\usepackage{bm}
\usepackage{verbatim}
\usepackage[utf8]{inputenc}
\usepackage{booktabs}
\usepackage{multirow}
\usepackage{subfig}
\usepackage{xcolor}
\usepackage[colorlinks=true,urlcolor=red,citecolor=red]{hyperref}
\usepackage[font=small]{caption}
\usepackage{float}
\usepackage{blindtext}

\newcommand{\mathsym}[1]{{}}

\input{tcilatex}
\begin{document}
\bibliographystyle{unsrt}
\title{Viable low-scale model with universal and inverse seesaw mechanisms}
\author{A. E. C\'{a}rcamo Hern\'{a}ndez}
\email{antonio.carcamo@usm.cl}
\author{Juan Marchant Gonz\'{a}lez}
\email{juan.marchantgonzalez@gmail.com}
\affiliation{Universidad T\'{e}cnica Federico Santa Mar\'{\i}a,\\
Centro Cient\'{\i}fico-Tecnol\'{o}gico de Valpara\'{\i}so,\\
Casilla 110-V, Valpara\'{\i}so, Chile}
\author{U. J. Salda\~na-Salazar}
\email{ulises.saldana@mpi-hd.mpg.de}
\affiliation{Max-Planck-Institut f\"ur Kernphysik,\\
Postfach 103980, D-69029 Heidelberg, Germany}
\date{\today }

\begin{abstract}
We formulate a viable low-scale seesaw model, where the masses for the
standard model (SM) charged fermions lighter than the top quark emerge from
a universal seesaw mechanism mediated by charged vectorlike fermions. The small light active neutrino masses are produced from an
inverse seesaw mechanism mediated by right-handed Majorana neutrinos. Our
model is based on the $A_{4} $ family symmetry, supplemented by cyclic
symmetries, whose spontaneous breaking produces the observed pattern of SM
fermion masses and mixings. The model can accommodate the muon and electron anomalous magnetic
dipole moments and predicts strongly suppressed $\mu\rightarrow e\gamma $ and $\tau \rightarrow \mu \gamma $ decay rates, but
allows a $\tau \rightarrow e\gamma $ decay within the reach of the
forthcoming experiments.
\end{abstract}

\maketitle

%%%%%%%%%%%%%%%%%%%%%%%%%

\section{Introduction}

%%%%%%%%%%%%%%%%%%%%%%%%% 
\noindent The standard model (SM) has offered us a theoretical framework
with great experimental success. In spite of this, the observed values in
quark mixing angles together with the pattern in the charged fermion masses
find no explanation. Moreover, the observation of neutrino oscillations has
augmented this puzzle as the theory must also be extended to incorporate
neutrino masses along with the observed leptonic mixing parameters. The
pattern in all the fermion masses may be described by three main aspects.
\begin{enumerate}
\item[(i)] Only one mass at the electroweak (EW) scale with all others
well below it, 
\begin{align}  \label{eq:MtEW}
m_t \sim \frac{v_\text{EW}}{\sqrt{2}} \gg
\{m_b,m_\tau,m_c,m_\mu,m_s,m_d,m_u,m_e\} \;.
\end{align}

\item[(ii)] Neutrino masses are much smaller than the electron mass, 
\begin{align}  \label{eq:mvSMALL}
m_\nu \lesssim \left( \frac{m_e}{m_t} \right) m_e\;.
\end{align}

\item[(iii)] The charged fermion masses satisfy a hierarchical structure, 
\begin{align}  \label{eq:massHier}
m_{f,3} \gg m_{f,2} \gg m_{f,1} \;, \qquad (f=u,d,e) \;.
\end{align}
\end{enumerate}

Several attempts have been made to theoretically describe each of these
aspects either individually~\cite%
{Minkowski:1977sc,Mohapatra:1979ia,Schechter:1980gr,Mohapatra:1986bd,Coy:2018bxr} or
various simultaneously; see for example~\cite%
{Froggatt:1978nt,Davidson:1987mh,Ibanez:1994ig,ArkaniHamed:1999dc,Kubo:2003iw,deMedeirosVarzielas:2006fc,Hernandez:2015hrt,CarcamoHernandez:2016pdu,Rodejohann:2019izm}%
. In the following, to produce Eqs.~\eqref{eq:MtEW} and~\eqref{eq:massHier}
we opt to work within a low-scale realization of a universal seesaw model for quarks whereas for the SM charged leptons the universal seesaw mechanism is supplemented by a Froggatt-Nielsen mechanism~\cite{Froggatt:1978nt}. In regards to the neutrino sector, to generate the small light active neutrino masses that satisfy Eq.~\eqref{eq:mvSMALL}, we consider an inverse seesaw mechanism. As it is shown in Sec. \ref{model}, 
despite the presence of several heavy vectorlike charged exotic leptons that trigger the universal seesaw mechanism, we assume that all of
them have masses of the same order of magnitude, thus implying the need of implementing a Froggatt-Nielsen
mechanism to generate the SM charged lepton mass hierarchy. Such a Froggatt-Nielsen mechanism is implemented by considering nonrenormalizable operators involving gauge singlet scalar fields charged under the discrete symmetries of the model, whose spontaneous breaking is crucial to yield the SM charged lepton mass hierarchy. Without those nonrenormalizable operators in the charged lepton sector, one can only explain the smallness of the tau lepton mass (in comparison with the electroweak scale), but one has to rely on an unnatural tuning in the charged lepton Yukawa couplings to address the hierarchy in the SM charged lepton masses.

In universal seesaw models~\cite{Davidson:1987mh,Deppisch:2017vne}, the
smallness of fermion masses except for the top quark, Eq.~\eqref{eq:MtEW},
can be easily explained by promoting parity symmetry (L$\leftrightarrow$R)
to a fundamental symmetry at high energies, larger than the Fermi scale.
These models are based on the $SU(2)_L \times SU(2)_R \times SU(3)_c \times
U(1)_{B-L}$ gauge symmetry, where $B$ and $L$ stand for the baryon and
lepton number, respectively. On the other hand, the matter content is
enlarged by introducing vector-like fermions (singlets under the left and
right isospin symmetries) whereas the scalar sector gets minimally enlarged
by mirroring the SM Higgs boson, $H\sim ({\ 2, 1,1,-1})$, to the right sector, $H_R
\sim (1,2,1,-1)$, transforming as a right doublet. The conventional bidoublet
in L-R symmetric theories is here missing. As a consequence neutrinos have
no masses and Yukawa interactions are now made with both scalars whereas the
singlet fermions acquire their own mass terms. Typically, after both scalars
have acquired their vacuum expectation values (VEV), small fermion masses
arise as an admixture of both VEVs and the heavy mass of the singlet
fermions, $m_f \sim v_\text{EW} (v_R/M_x)$, where $v_R \ll M_x$, while the
top-quark mass has no vectorlike fermion companion, and thus its mass is
simply given by the standard formula, $m_t \sim v_\text{EW}$. For last, the
hierarchy given in Eq.~\eqref{eq:massHier} may be understood by considering
exotic fermion masses with an inverse hierarchy, $M_{x1} \gg M_{x2} \gg
M_{x3}$. In the following, we mimic the main shared features among this
class of models and discuss a low-scale scenario.

The smallness of neutrino masses may have a different origin than that of
the charged fermions. Already their superlightness seems to point  to
this possibility. Hence, here we consider that two different mechanisms are
responsible for the observed patterns in the fermion masses. We choose to
study the mass nature of neutrinos via an inverse seesaw~\cite%
{Mohapatra:1986bd,Deppisch:2004fa,Deppisch:2005zm,Abada:2014vea}. 
This mechanism leads to
an effective mass parameter given by $m_\nu \sim (\tfrac{m_D}{M_E})^2 \mu$
where $m_D$ is the typical scale of a Dirac mass, $M_E$ the heavy scale of
the isosinglet leptons that conserve lepton number, and $\mu$ the mass
scale of the gauge singlet neutrinos responsible for breaking lepton number.
It follows that for small $\mu$ $m_\nu$ becomes small, which is
opposite to the standard seesaw, where the smallness of neutrino masses is
due to the largeness of the right-handed neutrino masses. The advantage of
using an inverse seesaw is that lepton flavor violation (LFV) rates do not
depend on the small magnitude of the lepton number violating scale, $\mu$,
while they vanish in standard seesaw scenarios.

In this work we propose a low-scale seesaw model with extended scalar and
fermion sectors, consistent with the current pattern of SM fermion masses
and mixings. In our model, the masses of the SM charged fermions lighter
than the top quark are generated from a universal seesaw mechanism mediated
by charged exotic vectorlike fermions. The small light active neutrino
masses arise from an inverse seesaw mechanism mediated by three sterile
neutrinos. In our model we use the $A_4$ family symmetry, which is
supplemented by other auxiliary symmetries, thus allowing one to have a viable
description of the current SM fermion mass spectrum and mixing parameters.
We have chosen the $A_4$ family symmetry since it is the smallest order
discrete group with one three-dimensional and three distinct one-dimensional
irreducible representations, where the three families of fermions can be
accommodated rather naturally. This group was used for the first time in
Ref. \cite{Ma:2001dn} and subsequently used in \cite%
{Babu:2002dz,Altarelli:2005yp,Altarelli:2005yx,He:2006dk,Rodejohann:2015hka,Karmakar:2016cvb,Borah:2017dmk,Chattopadhyay:2017zvs,CarcamoHernandez:2017kra,Ma:2017moj,CentellesChulia:2017koy,Bjorkeroth:2017tsz,Srivastava:2017sno,Belyaev:2018vkl,CarcamoHernandez:2018aon,Srivastava:2018ser,delaVega:2018cnx,Pramanick:2019qpg,CarcamoHernandez:2019kjy}
to provide a viable and predictive description of the SM fermion mass
spectrum and mixing parameters.

The outline for the rest of this paper is as follows. In Sec.~\ref{model}
we introduce the model, followed by discussions on the quark and lepton
masses and mixing in Secs.~\ref{sec:quarks} and~\ref{sec:leptonsector},
respectively. We devote Section~\ref{sec:pheno} to study some
phenomenological aspects of our model. Finally, in Sec.~\ref%
{sec:conclusions} we conclude.

%%%%%%%%%%%%%%%%%%%%%%%%%

\section{The model}

\label{model} %%%%%%%%%%%%%%%%%%%%%%%%%
\noindent Our model is an explicit realization of a tripermuting (TP)
scenario wherein the neutrino sector is transformed into the mass basis via
a mixing matrix of the form~\cite{Bazzocchi:2011ax}, 
\begin{equation}
|\mathbf{U}_{\nu }^{\text{TP}}|=\frac{1}{3}%
\begin{pmatrix}
2 & 2 & 1 \\ 
2 & 1 & 2 \\ 
1 & 2 & 2%
\end{pmatrix}%
\;.
\end{equation}%
In this type of scenario, the charged lepton sector must be built in such a
way that contributions arising from their mixing matrix, $\mathbf{U}_{e}$,
may help us to reproduce the experimentally observed values as the full
mixing matrix would then be given by $\mathbf{U}_{\ell }^{\text{th}}=\mathbf{%
U}_{e}\mathbf{U}_{\nu }^{\text{TP}}{}^{\dagger }$. In our model, the resulting 
leptonic mixing matrix corresponds to the experimentally observed deviation of the TP scenario.

In addition to the usual SM particle content, in order to implement the
universal seesaw mechanism producing the masses for the SM charged
fermions lighter than the top quark, we consider vectorlike quarks and
charged leptons, 
\begin{equation}
T_{i,L(R)}\sim (3,1)_{2/3}\;,\qquad B_{j,L(R)}\sim (3,1)_{-1/3}\;,\qquad
E_{k,L(R)}\sim (1,1)_{-1}\;,
\end{equation}%
where $i=1,2$, $j,k=1,2,3$ and their transformation assignments are given
under the SM gauge group, $\mathcal{G}_{\text{SM}}\ =SU(3)_{c}\times
SU(2)_{L}\times U(1)_{Y}$. Also, the scalar sector is appropriately
enlarged, apart from the Higgs doublet, $H$, with real scalar singlets
(flavons), 
\begin{align}
\text{quark sector: }& \qquad \{\chi _{1},\chi _{2},\Phi _{1k},\Phi
_{2k},\Phi _{3k}\} \\
\text{lepton sector: }& \qquad \{\sigma ,\eta _{1}\eta _{2},\rho _{1},\rho
_{2},\xi _{k},\zeta _{k},S_{k}\}
\end{align}%
where $k=1,2,3$ and we have classified the flavons into two different groups
depending in which sector they are relevant.

To control the arbitrariness in the Yukawa interactions we introduce $A_4$
as our flavor symmetry. $A_4$ has been found to be phenomenologically
interesting and successful in the neutrino sector~\cite%
{Altarelli:2005yx,King:2006np,Ma:2005qf}. Appendix~\ref{app:A4} has a brief
description of the group and multiplication rules. It is worth mentioning that the $A_4$ family symmetry is crucial to get viable mass matrix textures for the SM fermion sector, due to the fact that we are grouping the three left-handed leptons and the three right-handed SM down-type quarks into $A_4$ triplets, whereas the left-handed SM quarks, the right-handed SM up-type quarks and right-handed SM charged leptons are assigned in the different $A_4$ singlets, as shown in Tables~\ref%
{tab:modelquarks} and \ref{tab:modelleptons}. However,  
we have found that $A_4$ is insufficient to fully develop a predictive
theory. For this purpose, we employ 
the Abelian symmetry, $Z_2 \times Z_5$
with an additional $Z^\prime_2$ and $Z_4 \times Z_8$, in the quark and
lepton sectors, respectively. Such extra discrete cyclic symmetries allow one to reduce the number of model parameters and to treat the quark and lepton sectors independently, thus yielding a predictive framework consistent with the current pattern of SM fermion masses and mixings. In addition, in order to get a predictive model, one also needs to rely on specific VEV configurations for the $A_4$ triplets SM gauge singlet scalar fields. Notice, that thanks to the aforementioned cyclic symmetries, the gauge singlet scalar fields participating in the lepton Yukawa interactions are different than the ones appearing in the quark Yukawa terms.
 
 The field charge assignments under the group factors of the model are given in Tables~\ref%
{tab:modelquarks}-\ref{tab:modelscalars}.
\begin{table}[tbp]
\centering
\begin{tabular}{c|ccc|ccc|ccc|cccc|cccccc}
\toprule[0.13em] & $Q_{3L}$ & $Q_{2L}$ & $Q_{1L}$ & $u_{3R}$ & $u_{2R}$ & $%
u_{1R}$ & $d_{3R}$ & $d_{2R}$ & $d_{1R}$ & $T_{1L}$ & $T_{1R}$ & $T_{2L}$ & $%
T_{2R}$ & $B_{1L}$ & $B_{1R}$ & $B_{2L}$ & $B_{2R}$ & $B_{3L}$ & $B_{3R}$ \\ 
\hline
$\mathcal{G}_{SM}$ & \multicolumn{3}{c|}{$(\mathbf{3},\mathbf{2})_{1/6}$} & 
\multicolumn{3}{c|}{$(\mathbf{3},\mathbf{2})_{2/3}$} & \multicolumn{3}{c|}{$(%
\mathbf{3},\mathbf{2})_{-1/3}$} & \multicolumn{4}{c|}{$(\mathbf{3},\mathbf{1}%
)_{2/3}$} & \multicolumn{6}{c}{$(\mathbf{3},\mathbf{1})_{-1/3}$} \\ \hline
$A_4$ & $\mathbf{1}$ & $\mathbf{1^{\prime \prime }}$ & $\mathbf{1^{\prime }}$
& $\mathbf{1}$ & $\mathbf{1^{\prime }}$ & $\mathbf{1^{\prime }}$ & 
\multicolumn{3}{c|}{$\mathbf{3}$} & $\mathbf{1^{\prime }}$ & $\mathbf{%
1^{\prime }}$ & $\mathbf{1^{\prime \prime }}$ & $\mathbf{1^{\prime \prime }}$
& $\mathbf{1^{\prime }}$ & $\mathbf{1^{\prime }}$ & $\mathbf{1^{\prime
\prime }}$ & $\mathbf{1^{\prime \prime }}$ & $\mathbf{1}$ & $\mathbf{1}$ \\ 
$Z_2$ & $0$ & $0$ & $1$ & $0$ & $0$ & $1$ & \multicolumn{3}{c|}{0} & $1$ & $%
1 $ & $0$ & $0$ & $1$ & $1$ & $0$ & $0$ & $0$ & $0$ \\ 
$Z_5$ & $2$ & $2$ & $2$ & $4$ & $4$ & $4$ & \multicolumn{3}{c|}{0} & $4$ & $%
4 $ & $4$ & $4$ & $0$ & $0$ & $0$ & $0$ & $0$ & $0$ \\ 
$Z^{\prime }_2$ & $0$ & $0$ & $0$ & $0$ & $0$ & $1$ & \multicolumn{3}{c|}{0}
& $0$ & $0$ & $0$ & $0$ & $0$ & $0$ & $0$ & $0$ & $0$ & $0$ \\ 
$Z_4$ & $0$ & $0$ & $0$ & $0$ & $0$ & $0$ & \multicolumn{3}{c|}{0} & $0$ & $%
0 $ & $0$ & $0$ & $0$ & $0$ & $0$ & $0$ & $0$ & $0$ \\ 
$Z_8$ & $0$ & $0$ & $0$ & $0$ & $0$ & $0$ & \multicolumn{3}{c|}{0} & $0$ & $%
0 $ & $0$ & $0$ & $0$ & $0$ & $0$ & $0$ & $0$ & $0$ \\ 
\bottomrule[0.13em] % &  &  &  &  &  &  &  &  &  &  &  &  &  &  &  &  &  &  & 
\end{tabular}%
\caption{Fermion content in the quark sector and its charge assignment under
the discrete flavor symmetry group. }
\label{tab:modelquarks}
\end{table}

\begin{table}[tbp]
\begin{tabular}{c|ccc|ccc|ccc|ccc|cccccc}
\toprule[0.13em] & $\ell _{3L}$ & $\ell _{2L}$ & $\ell _{1L}$ & $e_{3R}$ & $%
e_{2R}$ & $e_{1R}$ & $N_{1R}$ & $N_{2R}$ & $N_{3R}$ & $\Omega _{1R}$ & $%
\Omega _{2R}$ & $\Omega _{3R}$ & $E_{1L}$ & $E_{1R}$ & $E_{2L}$ & $E_{2R}$ & 
$E_{3L}$ & $E_{3R}$ \\ \hline
$\mathcal{G}_{SM}$ & \multicolumn{3}{c|}{$(\mathbf{1},\mathbf{2})_{-1/2}$} & 
\multicolumn{3}{c|}{$(\mathbf{1},\mathbf{1})_{-1}$} & \multicolumn{3}{c|}{$(%
\mathbf{1},\mathbf{1})_{0}$} & \multicolumn{3}{c|}{$(\mathbf{1},\mathbf{1}%
)_{0}$} & \multicolumn{6}{c}{$(\mathbf{1},\mathbf{1})_{-1}$} \\ \hline
$A_{4}$ & \multicolumn{3}{c|}{$\mathbf{3}$} & $\mathbf{1^{\prime \prime }}$
& $\mathbf{1^{\prime }}$ & $\mathbf{1}$ & \multicolumn{3}{c|}{$\mathbf{3}$}
& \multicolumn{3}{c|}{$\mathbf{3}$} & $\mathbf{1}$ & $\mathbf{1}$ & $\mathbf{%
1^{\prime }}$ & $\mathbf{1^{\prime }}$ & $\mathbf{1^{\prime \prime }}$ & $%
\mathbf{1^{\prime \prime }}$ \\ 
$Z_{2}$ & \multicolumn{3}{c|}{0} & $0$ & $0$ & $0$ & \multicolumn{3}{c|}{1}
& \multicolumn{3}{c|}{0} & $0$ & $0$ & $0$ & $0$ & $0$ & $0$ \\ 
$Z_{5}$ & \multicolumn{3}{c|}{$-2$} & $2$ & $2$ & $2$ & \multicolumn{3}{c|}{2
} & \multicolumn{3}{c|}{$2$} & $-1$ & $1$ & $-1$ & $1$ & $-1$ & $1$ \\ 
$Z_{2}^{\prime }$ & \multicolumn{3}{c|}{0} & $0$ & $0$ & $0$ & 
\multicolumn{3}{c|}{0} & \multicolumn{3}{c|}{0} & $0$ & $0$ & $0$ & $0$ & $0$
& $0$ \\ 
$Z_{4}$ & \multicolumn{3}{c|}{0} & $2$ & $0$ & $2$ & \multicolumn{3}{c|}{0}
& \multicolumn{3}{c|}{$-1$} & $2$ & $2$ & $0$ & $0$ & $2$ & $2$ \\ 
$Z_{8}$ & \multicolumn{3}{c|}{$-1$} & $-1$ & $-2$ & $-3$ & 
\multicolumn{3}{c|}{$-2$} & \multicolumn{3}{c|}{2} & $1$ & $1$ & $0$ & $0$ & 
$-1$ & $-1$ \\ 
\bottomrule[0.13em] % &  &  &  &  &  &  &  &  &  &  &  &  &  &  &  &  &  & 
\end{tabular}%
\caption{Fermion content in the lepton sector and its charge assignment
under the discrete flavor symmetry group. }
\label{tab:modelleptons}
\end{table}

\begin{table}[tp]
\begin{tabular}{c|c|ccccc|cccccccc}
\toprule[0.13em] & $H$ & $\chi _{1}$ & $\chi _{2}$ & $\Phi _{1}$ & $\Phi
_{2} $ & $\Phi _{3}$ & $\sigma $ & $\eta _{1}$ & $\eta _{2}$ & $\rho _{1}$ & 
$\rho _{2}$ & $\xi $ & $\zeta $ & $S$ \\ \hline
$\mathcal{G}_{SM}$ & $(\mathbf{1},\mathbf{2})_{1/2}$ & \multicolumn{13}{c}{$(%
\mathbf{1},\mathbf{1})_{0}$} \\ \hline
$A_{4}$ & $\mathbf{1}$ & $\mathbf{1}$ & $\mathbf{1^{\prime }}$ & $\mathbf{3}$
& $\mathbf{3}$ & $\mathbf{3}$ & $\mathbf{1}$ & $\mathbf{1}$ & $\mathbf{%
1^{\prime \prime }}$ & $\mathbf{1}$ & $\mathbf{1}$ & $\mathbf{3}$ & $\mathbf{%
3}$ & $\mathbf{3}$ \\ 
$Z_{2}$ & $0$ & $0$ & $0$ & $1$ & $0$ & $0$ & $0$ & $0$ & $0$ & $-1$ & $0$ & 
$0$ & $-1$ & $-1$ \\ 
$Z_{5}$ & $2$ & $0$ & $0$ & $0$ & $0$ & $0$ & $0$ & $0$ & $0$ & $-1$ & $3$ & 
$0$ & $-2$ & $-2$ \\ 
$Z_{2}^{\prime }$ & $0$ & $1$ & $0$ & $0$ & $0$ & $0$ & $0$ & $0$ & $0$ & $0$
& $0$ & $0$ & $0$ & $0$ \\ 
$Z_{4}$ & $0$ & $0$ & $0$ & $0$ & $0$ & $0$ & $0$ & $-2$ & $-1$ & $-1$ & $0$
& $0$ & $0$ & $0$ \\ 
$Z_{8}$ & $0$ & $0$ & $0$ & $0$ & $0$ & $0$ & $-1$ & $0$ & $0$ & $0$ & $0$ & 
$0$ & $-1$ & $-1$ \\ 
\bottomrule[0.13em] % &  &  &  &  &  &  &  &  &  &  &  &  &  & 
\end{tabular}%
\caption{Scalar content and its charge assignment under the discrete flavor
symmetry group. }
\label{tab:modelscalars}
\end{table}

The full Yukawa terms can be written as $\mathcal{L}_{Y}=\mathcal{L}_{Y}^{u}+%
\mathcal{L}_{Y}^{d}+\mathcal{L}_{Y}^{e}+\mathcal{L}_{Y}^{\nu }$ where for
the up-type quarks, {\small 
\begin{equation}
-\mathcal{L}_{Y}^{u}=y_{t}\overline{Q}_{3L}\widetilde{H}u_{3R}+y_{1}^{u}%
\overline{Q}_{1L}\widetilde{H}T_{1R}+y_{2}^{u}\overline{Q}_{2L}\widetilde{H}%
T_{2R}+y_{3}^{u}\overline{T}_{1L}\chi _{1}u_{1R}+y_{4}^{u}\overline{T}%
_{2L}\chi _{2}u_{2R}+M_{T1}\overline{T}_{1L}T_{1R}+M_{T2}\overline{T}%
_{2L}T_{2R}+\text{H.c.}\;,  \label{eq:Uquarks}
\end{equation}%
} down-type quarks, 
\begin{equation}
\begin{split}
-\mathcal{L}_{Y}^{d}=& \;y_{1}^{d}\overline{Q}_{1L}HB_{1R}+y_{2}^{d}%
\overline{Q}_{2L}HB_{2R}+y_{3}^{d}\overline{Q}_{3L}HB_{3R}+y_{4}^{d}%
\overline{B}_{1L}\Phi _{1}D_{R}+y_{5}^{d}\overline{B}_{2L}\Phi
_{2}D_{R}+y_{6}^{d}\overline{B}_{2L}\Phi _{3}D_{R} \\
\;& +y_{7}^{d}\overline{B}_{3L}\Phi _{2}D_{R}+y_{8}^{d}\overline{B}_{3L}\Phi
_{3}D_{R}+M_{B1}\overline{B}_{1L}B_{1R}+M_{B2}\overline{B}_{2L}B_{2R}+M_{B3}%
\overline{B}_{3L}B_{3R}+\text{H.c.}\;,
\end{split}
\label{eq:Dquarks}
\end{equation}%
charged leptons, 
\begin{eqnarray}
-\tciLaplace _{Y}^{e}&=&y_{1}^{\left( l\right) }\left( \overline{l}_{L}H\xi
\right) _{\mathbf{\mathbf{1}}}E_{1R}\frac{\sigma ^{2}\eta _{1}}{\Lambda ^{4}}%
+y_{2}^{\left( l\right) }\left( \overline{l}_{L}H\xi \right) _{\mathbf{1}%
^{\prime \prime }}E_{2R}\frac{\sigma }{\Lambda ^{2}}+y_{3}^{\left( l\right)
}\left( \overline{l}_{L}H\xi \right) _{\mathbf{1^{\prime }}}E_{3R}\frac{\eta
_{1}^{\ast }}{\Lambda ^{2}}+y_{4}^{\left( l\right) }\left( \overline{l}%
_{L}H\xi \right) _{\mathbf{1^{\prime }}}E_{1R}\frac{\sigma ^{2}\left( \eta
_{2}^{\ast }\right) ^{2}}{\Lambda ^{5}}  \label{eq:chargedLep} \\
&&+y_{5}^{\left( l\right) }\left( \overline{l}_{L}H\xi \right) _{\mathbf{%
\mathbf{1}}}E_{3R}\frac{\eta _{2}^{2}}{\Lambda ^{3}}+x_{1}^{\left( l\right) }%
\overline{E}_{1L}\rho _{2}^{\ast }e_{1R}\frac{\left( \sigma ^{\ast }\right)
^{4}}{\Lambda ^{4}}+x_{2}^{\left( l\right) }\overline{E}_{2L}\rho _{2}^{\ast
}e_{2R}\frac{\left( \sigma ^{\ast }\right) ^{2}}{\Lambda ^{2}}+x_{3}^{\left(
l\right) }\overline{E}_{3L}\rho _{2}^{\ast
}e_{3R}+\sum_{i=1}^{3}y_{i}^{\left( E\right) }\overline{E}_{iL}\rho
_{2}E_{iR}+\text{ H.c.}\;\notag ,
\label{Lyl}
 \end{eqnarray}
and neutrinos,
\begin{equation} 
\begin{split}
-\tciLaplace _{Y}^{\nu }& =y_{1}^{\left( \nu \right) }\left( \overline{l}_{L}%
\widetilde{H}N_{R}\right) _{\mathbf{3s}}\frac{\zeta ^{\ast }}{\Lambda }%
+y_{2}^{\left( \nu \right) }\left( \overline{l}_{L}\widetilde{H}N_{R}\right)
_{\mathbf{3a}}\frac{\zeta ^{\ast }}{\Lambda }+y_{3}^{\left( \nu \right)
}\left( \overline{l}_{L}\widetilde{H}N_{R}\right) _{\mathbf{3s}}\frac{%
S^{\ast }}{\Lambda }+y_{4}^{\left( \nu \right) }\left( \overline{l}_{L}%
\widetilde{H}N_{R}\right) _{\mathbf{3a}}\frac{S^{\ast }}{\Lambda } 
\label{eq:neutrinos} \\
& +y^{\left( N\right) }\left( \overline{N}_{R}\Omega _{R}^{C}\right) _{%
\mathbf{1}}\rho _{1}+y^{\left( \Omega \right) }\left( \overline{\Omega _{R}}%
\Omega _{R}^{C}\right) _{\mathbf{3s}}\rho _{2}^{\ast }\frac{\left( \rho
_{1}^{\ast }\right) ^{2}\sigma ^{4}\eta _{1}}{\Lambda ^{7}}+\text{ H.c.}\;,
\end{split}
\end{equation}%
being the dimensionless couplings in Eqs.~\eqref{eq:Uquarks}-\eqref{eq:neutrinos} $\mathcal{%
O}(1)$ parameters. \newline

We denote by $\langle \chi _{i}\rangle =v_{\chi _{i}} \;(i=1,2)$, and assume
the following VEV patterns for the $A_{4}$ triplet SM singlet scalars $\Phi
_{1,2,3}$, $\xi $, $\zeta $ and $S$, 
\begin{align}
\langle \Phi _{1}\rangle & =\frac{v_{1}}{\sqrt{2}}(0,1,1)\;,\qquad \langle
\Phi _{2}\rangle =\frac{v_{2}}{\sqrt{3}}(1,1,1)\;, \qquad \langle \Phi
_{3}\rangle =v_{3}(0,0,1) \;, \\
\left\langle \xi \right\rangle & =\frac{v_{\xi }}{\sqrt{3}}\left(
1,1,1\right) ,\qquad \left\langle \zeta \right\rangle =v_{\zeta }\left(
0,0,1\right) ,\qquad \left\langle S\right\rangle =v_{S}\left( 1,0,0\right) ,
\label{VEV}
\end{align}%
which are natural solutions of the scalar potential minimization equations
for a large region of the parameter space as shown in Refs.~\cite%
{Memenga:2013vc,He:2006dk,Lin:2008aj,Branco:2009by,Ivanov:2014doa,Pramanick:2017wry}%
. As the hierarchy among charged fermion masses and quark mixing angles emerges from the spontaneous breaking of the $A_{4}\times Z_{2}\times
Z_{5}\times Z_{2}^{\prime }\times Z_{4}\times Z_{8}$ discrete group, we set
the VEVs of the SM singlet scalar fields $\sigma $, $\xi _{i}$, $\zeta _{i}$
($i=1,2,3$)\ with respect to the Wolfenstein parameter $\lambda =0.225$ and
the model cutoff $\Lambda $, as follows 
\begin{equation}
\{v_{\rho _{1}},v_{\rho _{2}}\}\sim \mathcal{O}(1)\text{ TeV}\ll \{v_{\zeta
},v_{S},v_{\xi },v_{\sigma },v_{\eta _{1}},v_{\eta _{2}}\}\sim \lambda
\Lambda \;.  \label{eq:VEV}
\end{equation}
As it is shown in Sec. \ref{sec:leptonsector}, the aforementioned assumption allows one to explain the SM charged lepton mass hierarchy since it allows one to relate the SM charged lepton masses with different powers of the Wolfenstein parameter $\lambda =0.225$ times $\mathcal{O}(1)$ coefficients. It is worth mentioning that the model cutoff scale can be interpreted as the scale of the UV completion of the model, e.g. the masses of Froggatt-Nielsen messenger fields.
%%%%%%%%%%%%%%%%%%%%%%%%%

\section{Quark masses and mixings}

\label{sec:quarks} %%%%%%%%%%%%%%%%%%%%%%%%%
\noindent Because of the symmetry assignments, the top quark does not mix with
the exotic vectorlike quarks and thus its mass is simply given as in the SM
via its Yukawa interaction with the Higgs doublet, $m_{t}\sim y_{t}v_{\text{%
EW}}$. On the other hand, the vectorlike quarks do mix with all the others
SM quarks. Then, from Eqs.~\eqref{eq:Uquarks} and~\eqref{eq:Dquarks}, the
quark mass matrices take the form
\begin{equation}
\mathbf{\mathcal{M}}_{5\times 5}^{\text{up}}=%
\begin{pmatrix}
0_{3\times 3} & \mathbf{M}_\text{L}^{u} \\ 
\mathbf{M}_\text{$\chi$}^u & \mathbf{M}_{\text{T}}%
\end{pmatrix}%
\;\qquad \text{and}\qquad \;\mathbf{\mathcal{M}}_{6\times 6}^{\text{down}}=%
\begin{pmatrix}
0_{3\times 3} & \mathbf{M}_\text{L}^d \\ 
\mathbf{M}_\text{$\Phi$}^d & \mathbf{M}_{\text{B}}%
\end{pmatrix}%
\;,
\end{equation}%
where 
\begin{equation}
\mathbf{M}_\text{L}^u=\frac{v_{\text{EW}}}{\sqrt{2}}%
\begin{pmatrix}
y_{1}^{u} & 0 \\ 
0 & y_{2}^{u}%
\end{pmatrix}%
,\qquad \mathbf{M}_\text{$\chi$}^u=%
\begin{pmatrix}
y_{3}^{u}v_{\chi _{1}} & 0 \\ 
0 & y_{4}^{u}v_{\chi _{2}}%
\end{pmatrix}%
,\qquad \mathbf{M}_{\text{T}}\ =%
\begin{pmatrix}
M_{T1} & 0 \\ 
0 & M_{T2}%
\end{pmatrix}%
,
\end{equation}%
\begin{equation}
\mathbf{M}_\text{L}^d=\frac{v_{\text{EW}}}{\sqrt{2}}%
\begin{pmatrix}
y_{1}^{d} & 0 & 0 \\ 
0 & y_{2}^{d} & 0 \\ 
0 & 0 & y_{3}^{d}%
\end{pmatrix}%
,\qquad \mathbf{M}_{\Phi}^d=%
\begin{pmatrix}
0 & y_{4}^{d}\frac{v_{1}}{\sqrt{2}} & \omega y_{4}^{d}\frac{v_{1}}{\sqrt{2}}
\\ 
\omega y_{5}^{d}\frac{v_{2}}{\sqrt{3}} & \omega y_{5}^{d}\frac{v_{2}}{\sqrt{3%
}} & \omega ^{2}(y_{5}^{d}\frac{v_{2}}{\sqrt{3}}+y_{6}^{d}v_{3}) \\ 
y_{7}^{d}\frac{v_{2}}{\sqrt{3}} & y_{7}^{d}\frac{v_{2}}{\sqrt{3}} & y_{7}^{d}%
\frac{v_{2}}{\sqrt{3}}+y_{8}^{d}v_{3}%
\end{pmatrix}%
,\qquad \mathbf{M}_{B}=%
\begin{pmatrix}
M_{B1} & 0 & 0 \\ 
0 & M_{B2} & 0 \\ 
0 & 0 & M_{B3}%
\end{pmatrix}%
.
\end{equation}

As the masses of the vectorlike quarks are much larger than the employed VEVs, $%
M_{T},M_{B}\gg \{v_{\text{EW}}, v_{\chi}, v_{i} \}$, the implementation of the universal seesaw
yields the following $3\times 3$ low-scale quark mass matrices, 
\begin{equation}
\mathbf{M}_{u}\simeq 
\begin{pmatrix}
m_{u} & 0 & 0 \\ 
0 & m_{c} & 0 \\ 
0 & 0 & m_{t}%
\end{pmatrix}%
\qquad \text{ and }\qquad \mathbf{M}_{d}\simeq 
\begin{pmatrix}
0 & a_{d} & \omega a_{d} \\ 
\omega b_{d} & b_{d} & \omega ^{2}(b_{d}+c_{d}) \\ 
d_{d} & d_{d} & d_{d}+e_{d}%
\end{pmatrix}%
\;,
\end{equation}%
where the up-type quark masses are given by, 
\begin{equation} \label{eq:upmasses}
m_{u}=\frac{v_{\text{EW}}}{\sqrt{2}}\frac{\widetilde{y}_{1}^{u}v_{\chi _{1}}%
}{M_{T1}}\;,\qquad m_{c}=\frac{v_{\text{EW}}}{\sqrt{2}}\frac{\widetilde{y}%
_{2}^{u}v_{\chi _{2}}}{M_{T2}}\;,\qquad m_{t}=\frac{v_{\text{EW}}}{\sqrt{2}}%
y_{t}\;,
\end{equation}%
and we have defined the parameters, 
\begin{equation*}
a_{d}=\frac{v_{\text{EW}}}{\sqrt{2}}\frac{\widetilde{y}_{1}^{d}v_{1}}{\sqrt{2%
}M_{B1}}\;,\qquad b_{d}=\frac{v_{\text{EW}}}{\sqrt{2}}\frac{\widetilde{y}%
_{2}^{d}v_{2}}{\sqrt{3}M_{B2}}\;,\qquad c_{d}=\frac{v_{\text{EW}}}{\sqrt{2}}%
\frac{\widetilde{y}_{3}^{d}v_{3}}{M_{B2}}\;,\qquad d_{d}=\frac{v_{\text{EW}}%
}{\sqrt{2}}\frac{\widetilde{y}_{4}^{d}v_{2}}{\sqrt{3}M_{B3}}\;,\qquad e_{d}=%
\frac{v_{\text{EW}}}{\sqrt{2}}\frac{\widetilde{y}_{5}^{d}v_{3}}{M_{B3}}\;,
\end{equation*}%
with $\widetilde{y}_{k}^{f}$ denoting the product of two different Yukawa
couplings that can be merged into a single one as both are $\mathcal{O}(1)$
parameters. From Eq.~\eqref{eq:upmasses} and the known hierarchy in the
up-quark masses we can estimate the ratio among the
heavy masses and VEVs to be
\begin{equation}
\frac{v_{\chi _{1}}}{M_{T1}}\sim 10^{-5}\qquad \text{and}\qquad \frac{%
v_{\chi _{2}}}{M_{T2}}\sim 10^{-2}\;,
\end{equation}%
which if we assume a single heavy scale, $M_{T}\equiv \{M_{T1},M_{T2}\}$,
then $v_{\chi _{2}}\sim \lambda ^{3}M_{T}$ and $v_{\chi _{1}}\sim \lambda
^{7}M_{T}$. For example, for the range $M_{T}\sim \lbrack 1,10]\text{ TeV}$
one gets $v_{\chi _{2}}\sim \lbrack 10,100]\text{ GeV}$ and $v_{\chi
_{1}}\sim \lbrack 0.1,1]\text{ MeV}$. States associated to
these scalars could also have masses in the same range but not necessarily. 
There are also possibilities of attaining small enough VEVs while preserving large
masses. For example, introduction of an additional singlet scalar could induce, via a soft-breaking term, small VEVs while keeping the initial large masses of the scalar fields. This point, however, is left aside as it would require a thorough study of the scalar potential which is beyond the scope of this work.\newline

In total we have five complex parameters, that is, ten free parameters to fit seven
observables. 
It can be shown that after redefining the phases of the left- and right-handed
quarks the number of free parameters reduces to seven.
A judicious choice of phases, for example, 
\begin{align}
Q_{1L} \rightarrow e^{ \frac{i}{2} \left( 2\text{arg}( a_d) + {\text{arg}(b_d) - \text{arg}(c_d)} \right) } Q_{1L}  \;, \qquad
Q_{2L} \rightarrow e^{\frac{i}{2} \left( {\text{arg}(b_d) + \text{arg}(c_d)} \right) } Q_{2L} \;,
\qquad
Q_{3L} \rightarrow e^{ i  {\text{ arg}(d_d)} } Q_{3L}  \; ,
\end{align}
may allow us to set the independent phases such that their effect is similar to having 
assumed in the initial matrix, 
\begin{equation}
\text{arg}(a_d)=-\text{arg}(b_d)=\text{arg}(c_d)\qquad \text{and}%
\qquad \text{arg}(d_d)=0\;.
\end{equation}   %
Hence, the set of independent parameters becomes 
\begin{equation}
\{|a_{d}|,\;|b_{d}|,\;|c_{d}|,\;|d_{d}|,\;|e_{d}|,\;\text{arg}(a_{d}),\;%
\text{arg}(e_{d})\}\;.
\end{equation}

\begin{table}[tp]
\begin{tabular}{c|ccccccc}
\toprule[0.13em] Observable & $m_b$ [GeV] & $m_s$ [GeV] & $m_d$ [GeV] & $|%
\mathbf{V}_{12}|$ & $|\mathbf{V}_{23}|$ & $|\mathbf{V}_{13}|$ & ${J}_{q}$ [$%
\times10^{-5}$] \\ \hline
Experimental Value & $2.86\pm0.02$ & $0.055^{+0.004}_{-0.002}$ & $%
0.0027^{+0.0003}_{-0.0002}$ & $0.22452\pm 0.00044$ & $0.04214\pm0.00076$ & $%
0.00365 \pm 0.00012$ & $3.18\pm0.15$ \\ 
Fit & $2.86 $ & $0.052$ & $0.0028$ & $0.22457$ & $0.04232$ & $0.00376$ & $%
3.02$ \\ 
\bottomrule[0.13em] % &  &  &  &  &  &  & 
\end{tabular}%
\caption{The most recent values for the mixing parameters come from the
PDG-2018~\protect\cite{Tanabashi:2018oca}. The masses are taken at the $M_Z$
scale from~\protect\cite{Saldana-Salazar:2018jes}. Minimization of the $%
\protect\chi^2$ function leads to the best-fit values appearing in Eq.%
~\eqref{eq:BF-quarks} with a quality of fit given by $\protect\chi^2_\text{%
d.o.f.}=3.15/7$. The masses and mixing implied by the model fit perfectly at 
$1\protect\sigma$. }
\label{tab:quarks}
\end{table}

We then perform a numerical fit to the set of parameters. The experimental
input parameters are the three down-type quark masses, the magnitudes of the
three independent mixing matrix elements, and the Jarlskog invariant. The
masses are taken at the $M_{Z}$ scale with a symmetrized $1\sigma $ error
taken to be the larger one. The employed input parameters are summarized in
Table~\ref{tab:quarks}. To measure the quality of the fit we use the
function, 
\begin{equation}
\begin{split}
\chi ^{2}& =\frac{(m_{d}^{\text{th}}-m_{d}^{\text{exp}})^{2}}{\sigma _{d}^{2}%
}+\frac{(m_{s}^{\text{th}}-m_{s}^{\text{exp}})^{2}}{\sigma _{s}^{2}}+\frac{%
(m_{b}^{\text{th}}-m_{b}^{\text{exp}})^{2}}{\sigma _{b}^{2}}+\frac{(|\mathbf{%
V}_{12}^{\text{th}}|-|\mathbf{V}_{12}^{\text{ckm}}|)^{2}}{\sigma _{12}^{2}}
\\
& +\frac{(|\mathbf{V}_{23}^{\text{th}}|-|\mathbf{V}_{23}^{\text{ckm}}|)^{2}}{%
\sigma _{23}^{2}}+\frac{(|\mathbf{V}_{13}^{\text{th}}|-|\mathbf{V}_{13}^{%
\text{ckm}}|)^{2}}{\sigma _{13}^{2}}+\frac{(J_{q}^{\text{th}}-J_{q}^{\text{%
exp}})^{2}}{\sigma _{J}^{2}}\;.
\end{split}%
\end{equation}%
Its minimization leads to the best-fit values, 
\begin{equation}
\begin{split}
|a_{d}|& =0.0114225\text{ GeV},\quad |b_{d}|=0.0215709\text{ GeV},\quad
|c_{d}|=0.130513\text{ GeV},\quad |d_{d}|=0.765595\text{ GeV}, \\
|e_{d}|& =1.97927\text{ GeV},\quad \text{arg}(a_{d})=5.39151\text{ rad}%
\;,\quad \text{arg}(e_{d})=0.605986\text{ rad}\;,
\end{split}
\label{eq:BF-quarks}
\end{equation}%
implying the observed down-type quark masses and mixing shown in Table~\ref%
{tab:quarks}. Moreover, we find that the two smallest mixing angles are
correlated among them and also with the Jarlskog invariant; see Fig.~\ref%
{fig:QuarksCorrelations}. At last, notice that our model prefers small
values of the Jarlskog invariant compared to the latest fit from the PDG~%
\cite{Tanabashi:2018oca}. 

From the best-fit values, Eq.~\eqref{eq:BF-quarks}, we can estimate the
required ratio among the heavy masses and VEVs in order to reproduce the
observed mild hierarchy among the fitted parameters, 
\begin{equation}
\frac{v_{1}}{M_{B1}}\sim 10^{-4}\;,\quad \frac{v_{2}}{M_{B2}}\sim
10^{-4}\;,\quad \frac{v_{3}}{M_{B3}}\sim 10^{-2}\;,\quad \frac{v_{2}}{M_{B3}}%
\sim 10^{-2}\;,\quad \frac{v_{3}}{M_{B2}}\sim 10^{-3.5}\;,
\end{equation}%
in such a way that all Yukawa couplings may still remain as $\mathcal{O}(1)$
parameters. All these ratios can be rewritten in terms of the heaviest mass, 
\begin{equation}
M_{B}\equiv \{M_{B1},M_{B2}\}\;,\quad M_{B3}\sim \lambda ^{3}M_{B}\;,\quad
\{v_{1},v_{2},v_{3}\}\sim \lambda ^{6}M_{B}\;.
\end{equation}%
These relations imply, for example, for the range $M_{B}\sim \lbrack 100,1000]%
\text{ TeV}$, $M_{B3}\sim \lbrack 1,10]\text{ TeV}$ and $%
\{v_{1},v_{2},v_{3}\}\sim \lbrack 10,100]\text{ GeV}$. \newline

\begin{figure}[tbp]
\subfloat[]{
		\includegraphics[scale=0.45]{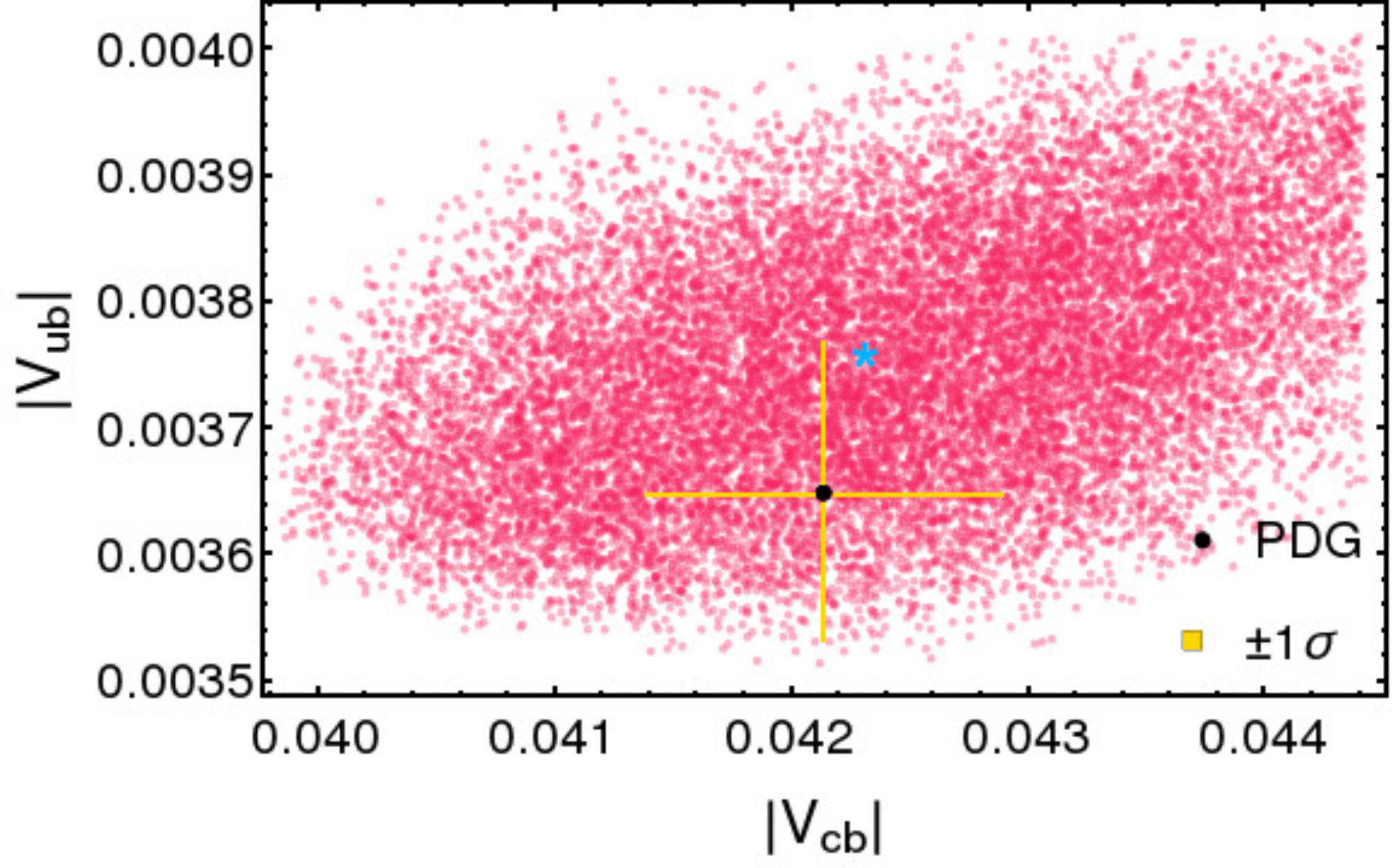}
		} \newline
\centering
\subfloat[]  {\  \includegraphics[scale=0.41]{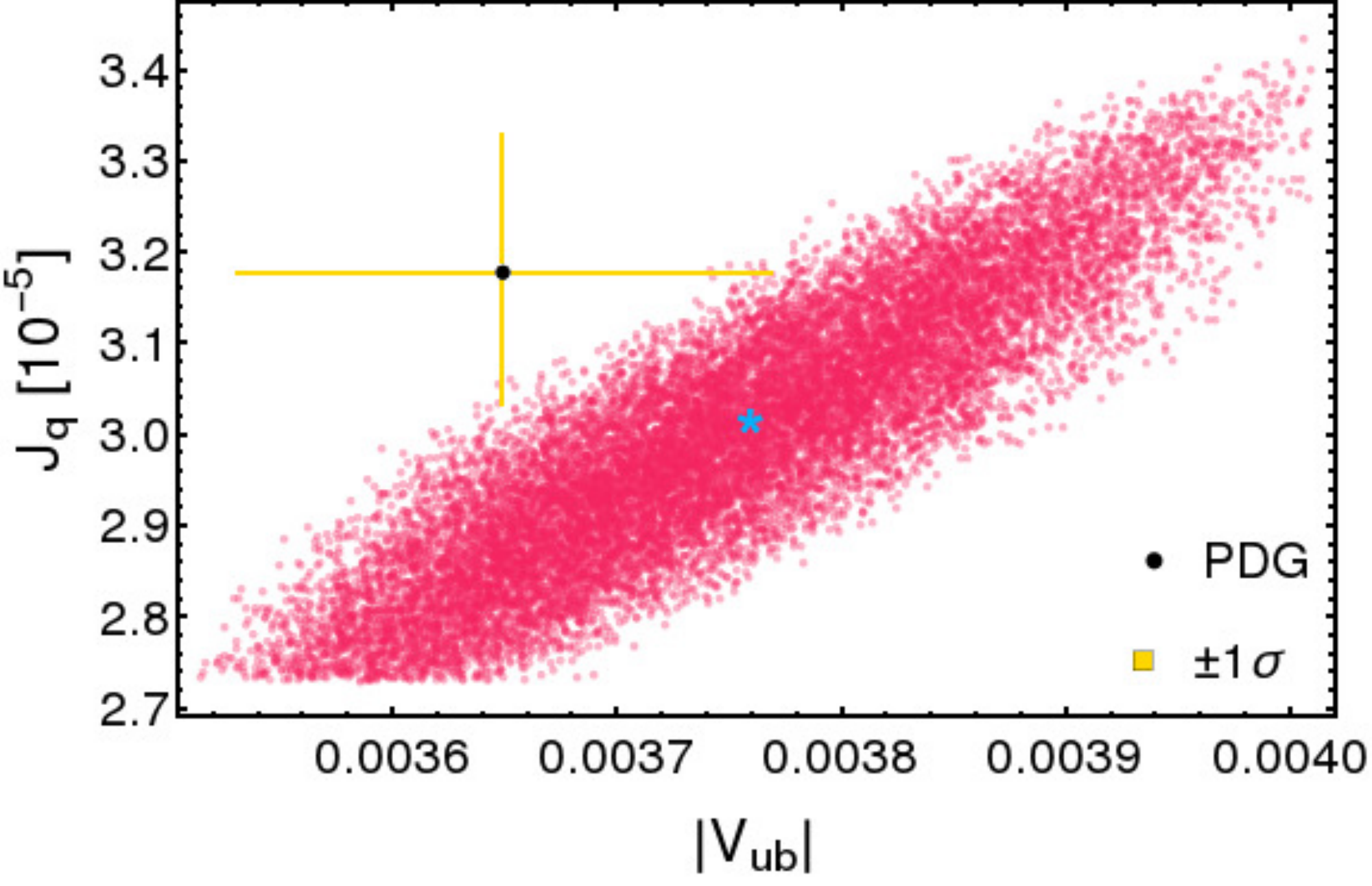}  } %
\subfloat[]  {\  \includegraphics[scale=0.4]{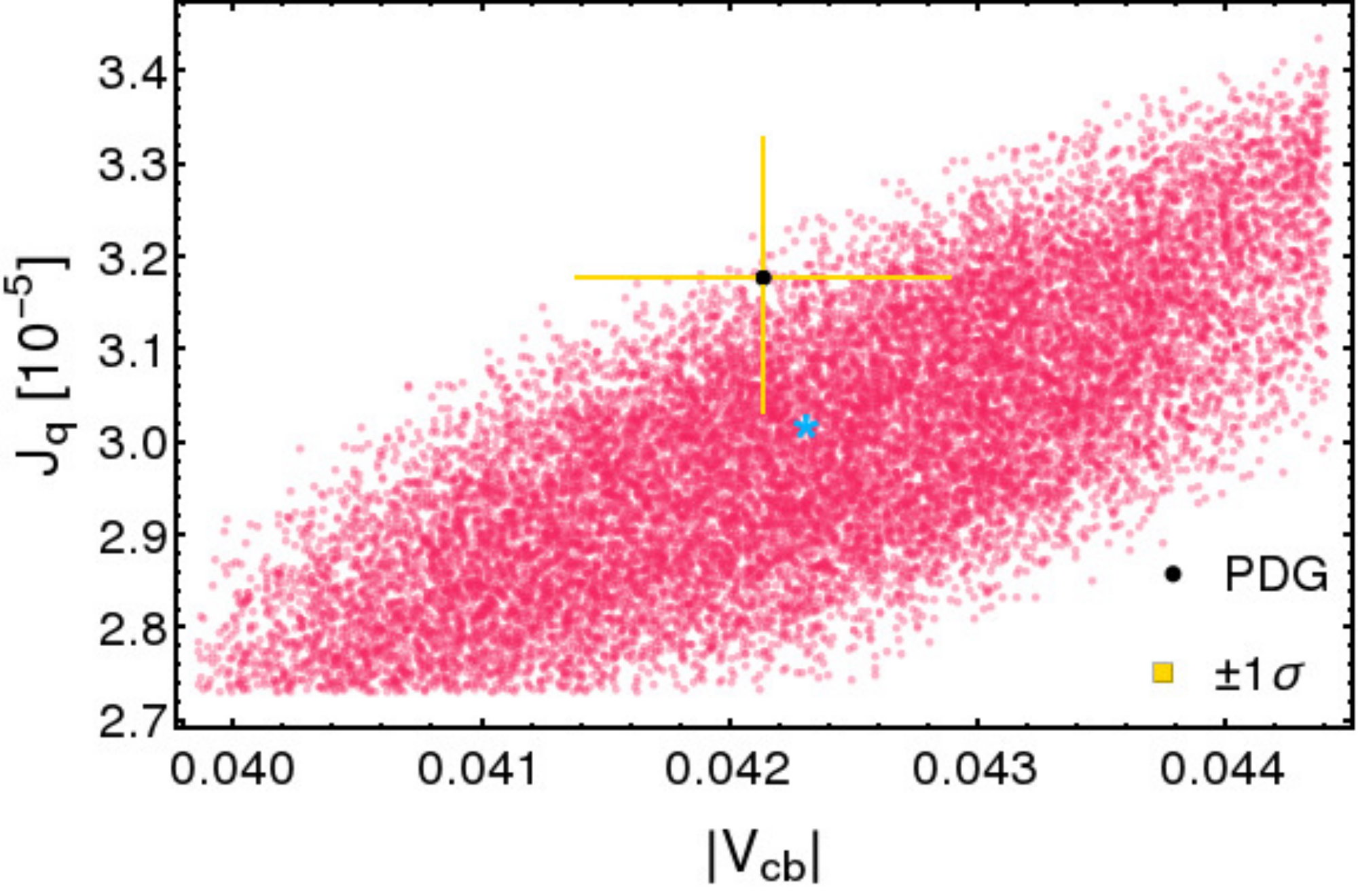}  }
\caption{Correlation plots between the two smallest quark mixing angles (top
panel) and between the Jarlskog invariant with each of them (bottom
panels). All the (red) points in the background are in agreement with the
experimental values within $3\protect\sigma$ deviations. The (blue) star
points represent the best-fit point of our minimization, whereas the black
points show the most recent values as taken from the PDG-2018 with their $1%
\protect\sigma$ deviations in (yellow) lines.}
\label{fig:QuarksCorrelations}
\end{figure}

Models with vectorlike fermions are being tested at the LHC. 
The ATLAS collaboration has reported several analyses, in particular~\cite{Aaboud:2017zfn,Aaboud:2018saj,Aaboud:2018pii}.  
At the moment, mass exclusion limits for exotic isosinglet quarks give
$M_B > 1.22$ TeV and $M_T > 1.31$ TeV as found in Ref.~\cite{Aaboud:2018pii}. 
These lower bounds were set by
only assuming that the exotic quarks can decay on SM particles. 
That is, the vectorlike quarks would first be produced at collider experiments
via pair production, a process dominated by the strong interactions, 
$gg\rightarrow \bar{B} B \, (\bar{T} T)$. Then, each exotic quark would decay 
to $T\rightarrow Wb, Zt, H t$ or $B\rightarrow Wt, Zb, Hb$, where it has only been
considered the third fermion family. Now, in our case, in the interaction basis,
our model has no initial mixing between the top-quark and the vector- and toplike fermions,
but only between the up and charm quarks with the exotic partners.
On the other hand, in the down-quark sector, all the standard quarks mix with the exotic
partners. Therefore, in the mass basis, we may only expect from the previous decay modes
that only those from the exotic bottomlike quarks will survive. 
Observation of an excess of events in any of the final states related to 
the vectorlike $B$ quarks with respect to the SM background, e.g., when
$Z\rightarrow {\ell}^+ \ell^-$ the final state has
one dilepton and at least two b-jets, would be a signal supporting this model at the LHC.
A detailed study of the collider phenomenology of our model is beyond the scope of this paper and is left for future studies.

%%%%%%%%%%%%%%%%%%%%%%%%%

\section{Lepton masses and mixings}

\label{sec:leptonsector} %%%%%%%%%%%%%%%%%%%%%%%%%
\noindent Using Eq. (\ref{Lyl}) we get the following mass matrix for charged leptons:
\begin{equation}
\mathbf{\mathcal{M}}_{6\times 6}^{\left( E\right) }=%
\begin{pmatrix}
0_{3\times 3} & \mathbf{M}_{1}^{\left( l\right) } \\ 
\mathbf{M}_{2}^{\left( l\right) } & \mathbf{M}_{E}%
\end{pmatrix}%
,  \label{ME}
\end{equation}%
where the different submatrices are given by 
\begin{eqnarray}
\mathbf{M}_{1}^{\left( l\right) } &=&\frac{v_\text{EW}v_{\xi }}{\sqrt{6}\Lambda }%
\left( 
\begin{array}{ccc}
1 & 1 & 1 \\ 
1 & \omega ^{2} & \omega \\ 
1 & \omega & \omega ^{2}%
\end{array}%
\right) \left( 
\begin{array}{ccc}
\cos \alpha & 0 & -e^{-i\gamma }\sin \alpha \\ 
0 & 1 & 0 \\ 
e^{i\gamma }\sin \alpha & 0 & \cos \alpha%
\end{array}%
\right) \left( 
\begin{array}{ccc}
y_{1}^{\left( l\right) }\lambda ^{2} & 0 & 0 \\ 
0 & y_{2}^{\left( l\right) }\lambda & 0 \\ 
0 & 0 & y_{3}^{\left( l\right) }\lambda%
\end{array}%
\right) ,  \notag \\
\mathbf{M}_{2}^{\left( l\right) } &=&\left( 
\begin{array}{ccc}
x_{1}^{\left( l\right) }\lambda ^{4} & 0 & 0 \\ 
0 & x_{2}^{\left( l\right) }\lambda ^{2} & 0 \\ 
0 & 0 & x_{3}^{\left( l\right) }%
\end{array}%
\right) v_{\rho _{2}},\qquad \qquad \mathbf{M}_{E}=\left( 
\begin{array}{ccc}
y_{1}^{\left( E\right) } & 0 & 0 \\ 
0 & y_{2}^{\left( E\right) } & 0 \\ 
0 & 0 & y_{3}^{\left( E\right) }%
\end{array}%
\right) v_{\rho _{2}}.  \label{MEblocks}
\end{eqnarray}
Here we have adopted a simplifying benchmark scenario with the following
particular assumptions about the charged lepton sector model parameters and
VEVs of some of the gauge singlet scalars 
\begin{equation}
y_{4}^{\left( l\right) }=e^{i\gamma }y_{1}^{\left( l\right) },\hspace{0.7cm}%
\hspace{0.7cm}y_{5}^{\left( l\right) }=-e^{-i\gamma }y_{3}^{\left( l\right)
},\hspace{0.7cm}\hspace{0.7cm}v_{\eta _{1}}=\lambda \cos \alpha \Lambda ,%
\hspace{0.7cm}\hspace{0.7cm}v_{\eta _{2}}=\sqrt{\lambda \sin \alpha }\Lambda
,\hspace{0.7cm}\hspace{0.7cm}v_{\sigma }=\lambda \Lambda .
\label{Benchmark}
\end{equation}
In what follows we limit ourselves to this scenario considering the relations  (\ref{Benchmark}) as additional constraints on our model parameter space. On the other hand it is possible that the relations (\ref{Benchmark}) arise in our model framework as a consequence of some unrecognized symmetry or can be attributed to a particular ultraviolet completion of the model. Notice that due to the fact that the Yukawa couplings are order 1 numbers, the first two assumptions regarding the magnitude may naturally follow without losing too much generality.
Thus, the universal seesaw mechanism gives rise to the following SM charged lepton mass matrix: 
\begin{eqnarray}
{\bf M}_{l} &=&\mathbf{M}_{1}^{\left( l\right) }\mathbf{M}_{E}^{-1}\mathbf{M}%
_{2}^{\left( l\right) }= {\bf R}_{lL} \text{diag}\left( m_{e},m_{\mu },m_{\tau }\right) ,%
\hspace{1cm}  \notag \\
{\bf R}_{lL} &=&\frac{1}{\sqrt{3}}\left( 
\begin{array}{ccc}
1 & 1 & 1 \\ 
1 & \omega ^{2} & \omega \\ 
1 & \omega & \omega ^{2}%
\end{array}%
\right) \left( 
\begin{array}{ccc}
\cos \alpha & 0 & -e^{-i\gamma }\sin \alpha \\ 
0 & 1 & 0 \\ 
e^{i\gamma }\sin \alpha & 0 & \cos \alpha%
\end{array}%
\right) ,\hspace{1cm}\omega =e^{\frac{2\pi i}{3}},  \label{Ml}
\end{eqnarray}%
where the charged lepton masses are
\begin{equation}
m_{e}=\frac{x_{1}^{\left( l\right) }y_{1}^{\left( l\right) }\lambda
^{6}v_\text{EW}v_{\xi }}{\sqrt{6}y_{1}^{\left( E\right) }\Lambda }=a_{1}^{\left(
l\right) }\lambda ^{9}\frac{v_\text{EW}}{\sqrt{2}},\hspace{0.5cm}m_{\mu }=\frac{%
x_{2}^{\left( l\right) }y_{2}^{\left( l\right) }\lambda ^{3}v_\text{EW}v_{\xi }}{%
\sqrt{6}y_{1}^{\left( E\right) }\Lambda }=a_{2}^{\left( l\right) }\lambda
^{5}\frac{v_\text{EW}}{\sqrt{2}},\hspace{0.5cm}m_{\tau }=\frac{x_{3}^{\left( l\right)
}y_{3}^{\left( l\right) }v_\text{EW}v_{\xi }}{\sqrt{6}y_{1}^{\left( E\right)
}\Lambda }=a_{3}^{\left( l\right) }\lambda ^{3}\frac{v_\text{EW}}{\sqrt{2}}.
\label{leptonmasses}
\end{equation}%
Here we have considered $x_{i}^{\left( l\right) }\sim y_{i}^{\left( l\right)
}\lesssim \mathcal{O}(1)$ ($i=1,2,3$) and $y_{i}^{\left( E\right) }\lesssim 
\mathcal{O}(\sqrt{4\pi })$. Let us note that the charged lepton masses are
linked with the scale of electroweak symmetry breaking through their power
dependence on the Wolfenstein parameter $\lambda =0.225$, with $\mathcal{O}%
(1)$ coefficients.

The neutrino Yukawa terms of Eq. \eqref{eq:neutrinos} give origin to
the following neutrino mass terms: 
\begin{equation}
-\mathcal{L}_{mass}^{\left( \nu \right) }=\frac{1}{2}\left( 
\begin{array}{ccc}
\overline{\nu _{L}^{C}} & \overline{N}_{R} & \overline{\Omega }_{R}%
\end{array}%
\right) {\bf M}_{\nu }\left( 
\begin{array}{c}
\nu _{L} \\ 
N_{R}^{C} \\ 
\Omega _{R}^{C}%
\end{array}%
\right) +H.c,  \label{Lnu}
\end{equation}%
where the neutrino mass matrix reads%
\begin{equation}
{\bf M}_{\nu }=\left( 
\begin{array}{ccc}
0_{3\times 3} & {\bf M}_{1} & 0_{3\times 3} \\ 
{\bf M}_{1}^{T} & 0_{3\times 3} & {\bf M}_{2} \\ 
0_{3\times 3} & {\bf M}_{2}^{T} & {\bf \mu}%
\end{array}%
\right)  \label{Mnu}
\end{equation}%
and the submatrices read 
\begin{eqnarray}
{\bf M}_{1} &=&\frac{v_{\zeta }v_\text{EW}}{\sqrt{2}\Lambda }\left( 
\begin{array}{ccc}
0 & y_{1}^{\left( \nu \right) }+y_{2}^{\left( \nu \right) } & 0 \\ 
y_{1}^{\left( \nu \right) }-y_{2}^{\left( \nu \right) } & 0 & r\left(
y_{1}^{\left( \nu \right) }+y_{2}^{\left( \nu \right) }\right) \\ 
0 & r\left( y_{1}^{\left( \nu \right) }-y_{2}^{\left( \nu \right) }\right) & 
0%
\end{array}%
\right) \\
&=&f\left( 
\begin{array}{ccc}
0 & x & 0 \\ 
y & 0 & rx \\ 
0 & ry & 0%
\end{array}%
\right) ,\hspace{0.7cm}f=\frac{v_{\zeta }v_\text{EW}}{\sqrt{2}\Lambda },\hspace{0.7cm}%
r=\frac{v_{S}}{v_{\zeta }},\hspace{0.7cm}x=y_{1}^{\left( \nu \right)
}+y_{2}^{\left( \nu \right) },\hspace{0.7cm}y=y_{1}^{\left( \nu \right)
}-y_{2}^{\left( \nu \right) }, \\
{\bf M}_{2} &=&m_{N}\left( 
\begin{array}{ccc}
1 & 0 & 0 \\ 
0 & 1 & 0 \\ 
0 & 0 & 1%
\end{array}%
\right) ,\hspace{0.7cm}\hspace{0.7cm}{\bf \mu} =\frac{y^{\left( \Omega \right)
}v_{\rho _{2}}v_{\rho _{1}}^{2}v_{\sigma }^{4}v_{\eta _{1}}}{\Lambda ^{7}}%
\left( 
\begin{array}{ccc}
0 & 1 & -1 \\ 
1 & 0 & 1 \\ 
-1 & 1 & 0%
\end{array}%
\right) ,\hspace{0.7cm}\hspace{0.7cm}m_{N}=y^{\left( N\right) }v_{\rho _{1}}.
\end{eqnarray}%
The light active masses arise from an inverse seesaw mechanism and the
resulting physical neutrino mass matrices take the form 
\begin{equation}
\widetilde{\bf M}_{\nu }={\bf M}_{1}\left( {\bf M}_{2}^{T}\right) ^{-1} {\bf \mu}
{\bf M}_{2}^{-1}{\bf M}_{1}^{T},\hspace{0.7cm}{\bf M}_{\nu }^{\left( 1\right) }=-\frac{1}{2}%
\left( {\bf M}_{2}+{\bf M}_{2}^{T}\right) +\frac{1}{2}{\bf \mu} ,\hspace{0.7cm}{\bf M}_{\nu
}^{\left( 2\right) }=\frac{1}{2}\left( {\bf M}_{2}+{\bf M}_{2}^{T}\right) +\frac{1}{2}%
{\bf \mu} ,  \label{M1nu}
\end{equation}%
where $\widetilde{\bf M}_{\nu }$ corresponds to the active neutrino mass matrix
whereas ${\bf M}_{\nu }^{\left( 1\right) }$ and ${\bf M}_{\nu }^{\left( 2\right) }$ are
the sterile mass matrices.

Thus, the light active neutrino mass matrix is given by 
\begin{eqnarray}
\widetilde{\bf M}_{\nu } &=&\frac{y^{\left( \Omega \right) }v_{\rho _{2}}v_{\rho
_{1}}^{2}v_{\sigma }^{4}v_{\eta _{1}}f^{2}}{m_{N}^{2}\Lambda ^{7}}\left( 
\begin{array}{ccc}
0 & x\left( y+rx\right) & 0 \\ 
x\left( y+rx\right) & -2rxy & ry\left( y+rx\right) \\ 
0 & ry\left( y+rx\right) & 0%
\end{array}%
\right) \allowbreak \\
&=&m_{\nu }\left( 
\begin{array}{ccc}
0 & x\left( y+rx\right) & 0 \\ 
x\left( y+rx\right) & -2rxy & ry\left( y+rx\right) \\ 
0 & ry\left( y+rx\right) & 0%
\end{array}%
\right) ,\hspace{0.7cm}\hspace{0.7cm}\allowbreak m_{\nu }=\frac{y^{\left(
\Omega \right) }v_{\rho _{2}}v_{\rho _{1}}^{2}v_{\sigma }^{4}v_{\eta
_{1}}f^{2}}{m_{N}^{2}\Lambda ^{7}}.
\end{eqnarray}

The full neutrino mass matrix given by Eq. (\ref{Mnu}) can be diagonalized by the following rotation matrix \cite{Catano:2012kw}:
\begin{equation}
\mathbb{{R}}=%
\begin{pmatrix}
{\bf R}_{\nu } & {\bf R}_{1}{\bf R}_{M}^{\left( 1\right) } & {\bf R}_{2}{\bf R}_{M}^{\left( 2\right) } \\ 
-\frac{({\bf R}_{1}^{\dagger }+{\bf R}_{2}^{\dagger })}{\sqrt{2}}{\bf R}_{\nu } & \frac{(1-{\bf S})}{%
\sqrt{2}}{\bf R}_{M}^{\left( 1\right) } & \frac{(1+{\bf S})}{\sqrt{2}}{\bf R}_{M}^{\left(
2\right) } \\ 
-\frac{({\bf R}_{1}^{\dagger }-{\bf R}_{2}^{\dagger })}{\sqrt{2}}{\bf R}_{\nu } & \frac{(-1-{\bf S})%
}{\sqrt{2}}{\bf R}_{M}^{\left( 1\right) } & \frac{(1-{\bf S})}{\sqrt{2}}{\bf R}_{M}^{\left(
2\right) }%
\end{pmatrix}%
,  \label{U}
\end{equation}%
where 
\begin{equation}
{\bf S}=-\frac{1}{4}{\bf M}_{2}^{-1}{\bf \mu} ,\hspace{1cm}\hspace{1cm}{\bf R}_{1}\simeq {\bf R}_{2}\simeq 
\frac{1}{\sqrt{2}}{\bf M}_{1}^{\ast }{\bf M}_{2}^{-1}.
\end{equation}

Notice that the physical neutrino spectrum is composed of three light active
neutrinos and six exotic neutrinos. The exotic neutrinos are pseudo-Dirac,
with masses $\sim \pm \frac{1}{2}\left( {\bf M}_{2}+{\bf M}_{2}^{T}\right) $ and a small
splitting ${\bf \mu} $. Furthermore, ${\bf R}_{\nu }$, ${\bf R}_{M}^{\left( 1\right) }$, and $%
{\bf R}_{M}^{\left( 2\right) }$ are the rotation matrices that diagonalize $%
\widetilde{{\bf M}}_{\nu }$, ${\bf M}_{\nu }^{\left( 1\right) }$, and ${\bf M}_{\nu }^{\left(
2\right) }$, respectively.

On the other hand, using Eq. (\ref{U}) we find that the neutrino fields $\nu
_{L}=\left( \nu _{1L},\nu _{2L},\nu _{3L}\right) ^{T}$, $\nu _{R}^{C}=\left(
\nu _{1R}^{C},\nu _{2R}^{C}\right) $, and $N_{R}^{C}=\left(
N_{1R}^{C},N_{2R}^{C}\right) $ are related with the physical neutrino fields by the following relations: 
\begin{equation}
\left( 
\begin{array}{c}
\nu _{L} \\ 
\nu _{R}^{C} \\ 
N_{R}^{C}%
\end{array}%
\right) =\mathbb{R}\Omega _{L}\simeq 
\begin{pmatrix}
{\bf R}_{\nu } & {\bf R}_{1}{\bf R}_{M}^{\left( 1\right) } & {\bf R}_{2}{\bf R}_{M}^{\left( 2\right) } \\ 
-\frac{({\bf R}_{1}^{\dagger }+{\bf R}_{2}^{\dagger })}{\sqrt{2}}{\bf R}_{\nu } & \frac{(1-{\bf S})}{%
\sqrt{2}}{\bf R}_{M}^{\left( 1\right) } & \frac{(1+{\bf S})}{\sqrt{2}}{\bf R}_{M}^{\left(
2\right) } \\ 
-\frac{({\bf R}_{1}^{\dagger }-{\bf R}_{2}^{\dagger })}{\sqrt{2}}{\bf R}_{\nu } & \frac{(-1-{\bf S})%
}{\sqrt{2}}{\bf R}_{M}^{\left( 1\right) } & \frac{(1-{\bf S})}{\sqrt{2}}{\bf R}_{M}^{\left(
2\right) }%
\end{pmatrix}%
\left( 
\begin{array}{c}
\Psi _{L}^{\left( 1\right) } \\ 
\Psi _{L}^{\left( 2\right) } \\ 
\Psi _{L}^{\left( 3\right) }%
\end{array}%
\right) ,\hspace{0.5cm}\hspace{0.5cm}\hspace{0.5cm}\hspace{0.5cm}\Psi
_{L}=\left( 
\begin{array}{c}
\Psi _{L}^{\left( 1\right) } \\ 
\Psi _{L}^{\left( 2\right) } \\ 
\Psi _{L}^{\left( 3\right) }%
\end{array}%
\right) ,
\end{equation}%
where $\Psi _{jL}^{\left( 1\right) }$, $\Psi _{jL}^{\left( 2\right) }$ and $%
\Psi _{jL}^{\left( 3\right) }$ ($j=1,2,3$) are the three active neutrinos
and six exotic neutrinos, respectively.

By varying the lepton sector model parameters, we find values for the
neutrino mass squared splittings, i.e, $\Delta m_{21}^{2}$ and $\Delta
m_{31}^{2}$, leptonic mixing angles $\theta^{(l)}_{12}$, $\theta^{(l)}_{23}$%
, and $\theta^{(l)}_{13}$, and the Dirac leptonic $CP$ violating phase consistent
with the neutrino oscillation experimental data, as indicated in Table \ref%
{neutrinos}. 

\begin{table}[tp]
\begin{tabular}{c|c|cccccc}
\toprule[0.13em] Observable & Range & $\Delta m_{21}^{2}$ [$10^{-5}$eV$^{2}$] & $\Delta m_{31}^{2}$ [$10^{-3}$eV$^{2}$] & $\theta^{(l)}_{12} (^{\circ })$ & $\theta^{(l)}_{13} (^{\circ })$ & $\theta^{(l)}_{23} (^{\circ })$ & $\delta^{(l)}_{CP} (^{\circ })$ \\ \hline
Experimental & $1\sigma$ & $%
7.39_{-0.20}^{+0.21}$ & $2.525_{-0.032}^{+0.033}$ & $33.82_{-0.76}^{+0.78}$ & $8.61_{-0.13}^{+0.13}$ & $49.6_{-1.2}^{+1.0}$ & $215_{-29}^{+40}$ \\
Value from Ref. \cite{deSalas:2017kay} & $3\sigma$ & $6.79-8.01$ & $2.427-2.625$ & $31.61-36.27 $ & $8.22-8.99$ & $40.3-52.4$ & $125-392$ \\
\cline{2-8}
Experimental & $1\sigma$ & $%
7.55_{-0.16}^{+0.20}$ & $2.50\pm 0.03$ & $34.5_{-1.0}^{+1.2}$ & $8.45_{-0.14}^{+0.16}$ & $47.7_{-1.7}^{+1.2}$ & $218_{-27}^{+38}$ \\ 
Value from Ref. \cite{Esteban:2018azc}& $3\sigma$ & $7.05-8.14$ & $2.41-2.60$ & $31.5-38.0 $ & $8.0-8.9$ & $41.8-50.7$ & $157-349$ \\
\cline{2-8}
Fit &$1\sigma$& $7.55$ & $2.50$ & $34.45$ & $8.45$ & $43.1$ & $218.2$ \\ 
\bottomrule[0.13em]% &  &  &  &  &  &  
\end{tabular}%
\caption{Model predictions for the scenario of normal neutrino mass hierarchy. The experimental values are taken from Refs. 
\protect\cite{deSalas:2017kay,Esteban:2016qun}.}
\label{neutrinos}
\end{table}

\begin{figure}[t]
\includegraphics[width=0.55\textwidth]{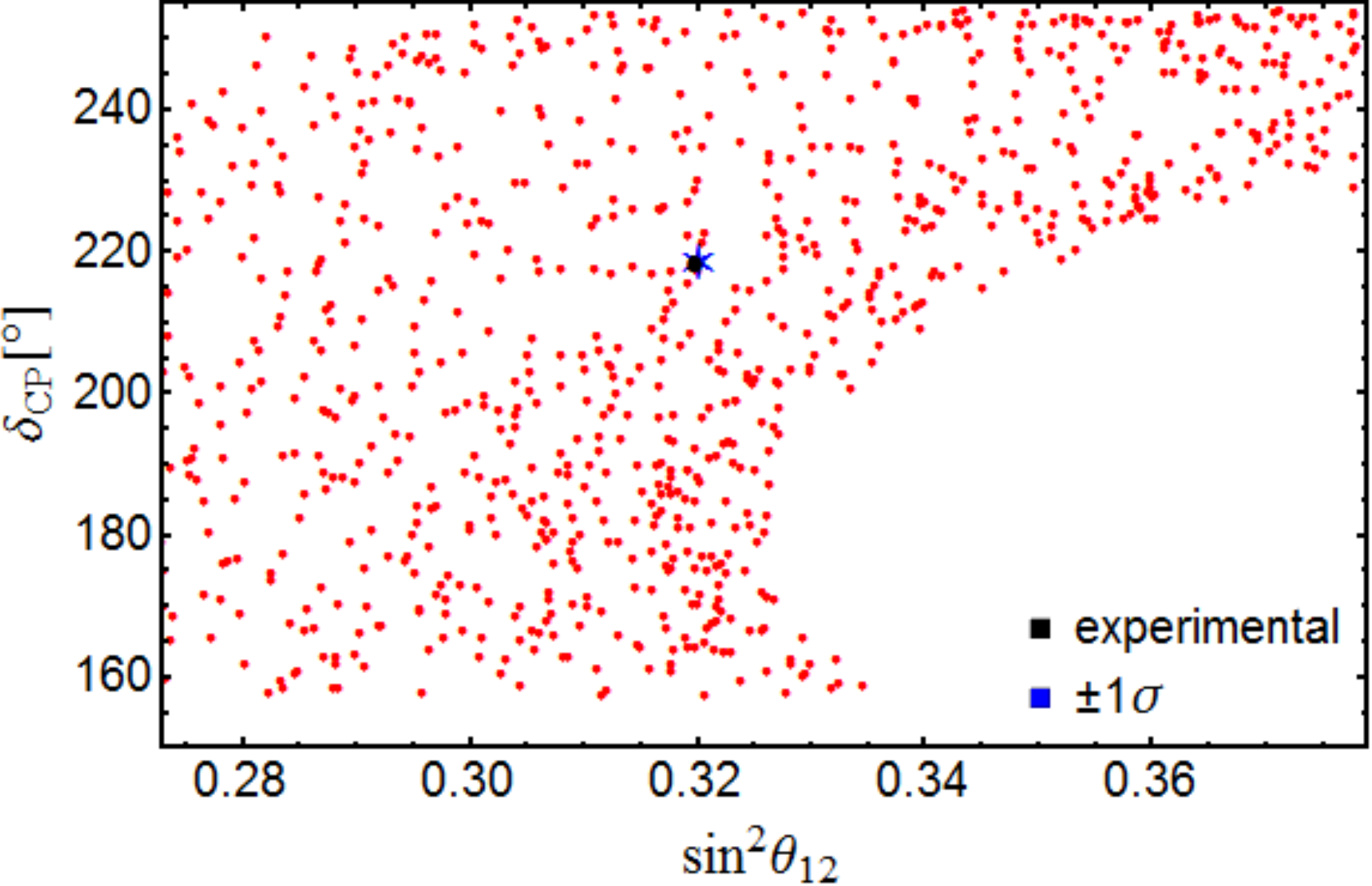}
\caption{Correlation between the solar mixing parameter $\sin ^{2}\protect%
\theta _{12}$ and the leptonic $CP$ violating phase.}
\label{Correlations12del}
\end{figure}

Fig.~\ref{Correlations12del} shows the correlation between the solar mixing
parameter $\sin ^{2}\theta _{12}$ and the leptonic $CP$ violating phase. To
obtain this figure the lepton sector model parameters were randomly
generated in a range of values where the neutrino mass squared splittings
and leptonic mixing parameters are inside the $3\sigma $ experimentally
allowed range. As seen from Fig.~\ref{Correlations12del}, our model
predicts a solar mixing parameter $\sin ^{2}\theta _{12}$ and leptonic Dirac
$CP$ violating phase in the ranges $0.27$ $\lesssim \sin ^{2}\theta
_{12}\lesssim 0.38$ and $140^{\circ }\lesssim \delta \lesssim 260^{\circ }$,
respectively. 

Another relevant observable that can be determined in this model  is the
effective Majorana neutrino mass parameter of neutrinoless double beta
decay, which provides information on the Majorana nature of neutrinos. The
effective Majorana neutrino mass parameter takes the form
\begin{equation}
m_{ee}=\left\vert \sum_{j}U_{ek}^{2}m_{\nu _{k}}\right\vert ,  \label{mee}
\end{equation}%
where $U_{ej}$ and $m_{\nu _{k}}$ are the the Pontecorvo-Maki-Nakagawa-Sakata
leptonic mixing matrix
elements and the neutrino Majorana masses, respectively. The neutrinoless
double beta ($0\nu \beta \beta $) decay amplitude is proportional to $m_{ee}$%
. 
\begin{figure}[tbp]
\centering
\includegraphics[width=0.55\textwidth]{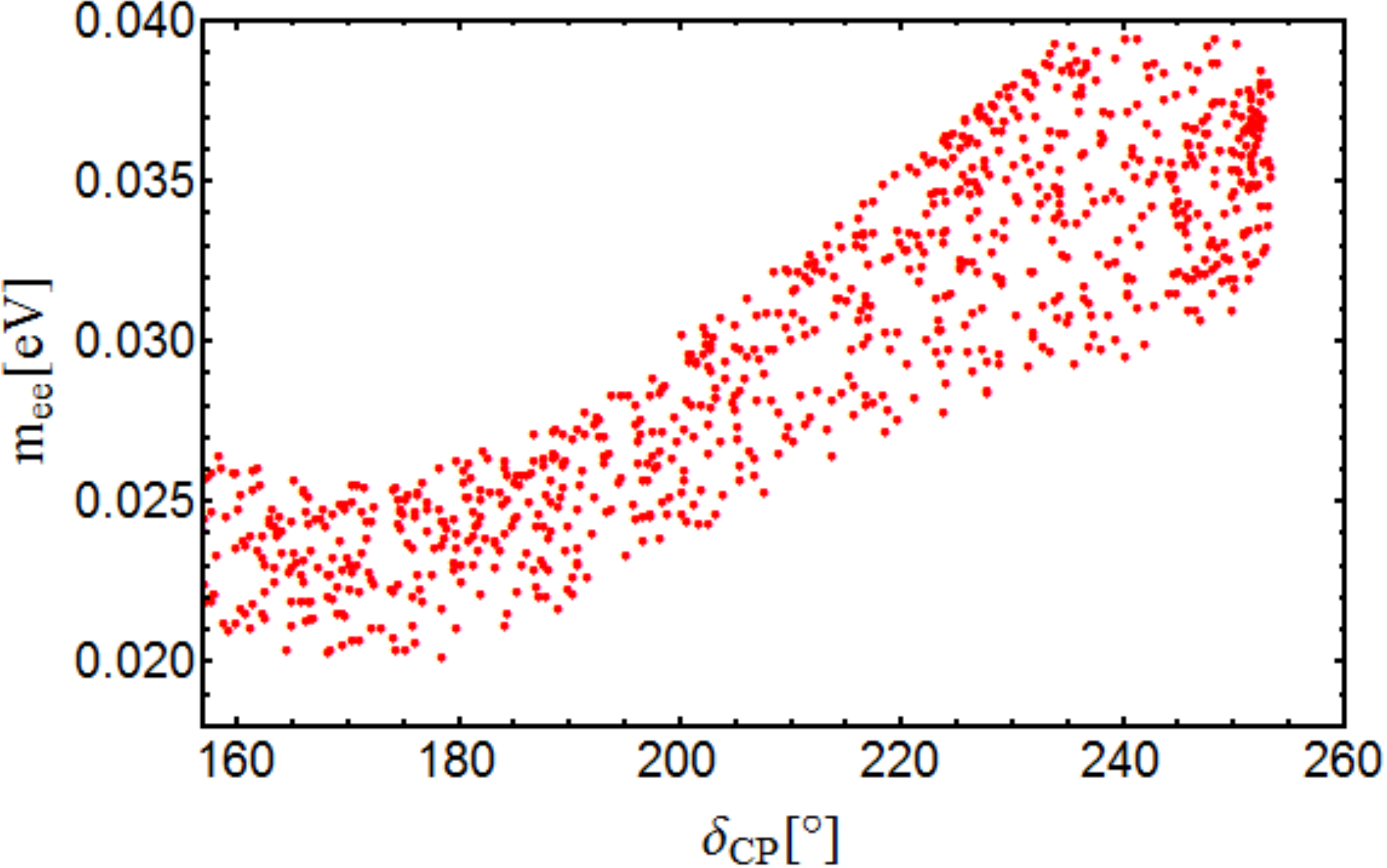}\newline
\caption{Correlation of the effective Majorana neutrino mass parameter $%
m_{ee}$ with the leptonic Dirac $CP$ violating phase $\protect\delta _{CP}$.}
\label{Correlationmee}
\end{figure}
In Fig.~\ref{Correlationmee} we display the correlation between the
effective Majorana neutrino mass parameter $m_{ee}$ and the leptonic Dirac
$CP$ violating phase $\delta _{CP}$. As indicated by Fig. \ref%
{Correlationmee}, our model predicts an effective Majorana neutrino mass
parameter in the range $0.020$ eV$\lesssim m_{ee}\lesssim $ $0.040$ eV, thus
implying that the values for the effective Majorana neutrino mass parameter
predicted by our model are within the reach of the next-generation
bolometric CUORE experiment \cite{Alduino:2017pni}, as well as the
next-to-next-generation ton-scale $0\nu \beta \beta $-decay experiments \cite%
{KamLAND-Zen:2016pfg,Albert:2017owj,Abt:2004yk,Gilliss:2018lke}. The current
most stringent experimental upper bound on the effective Majorana neutrino
mass parameter, i.e., $m_{ee}\leq 160$ meV, arises from the KamLAND-Zen limit
on the ${}^{136}\mathrm{Xe}$ $0\nu \beta \beta $ decay half-life $%
T_{1/2}^{0\nu \beta \beta }({}^{136}\mathrm{Xe})\geq 1.07\times 10^{26}$ yr 
\cite{KamLAND-Zen:2016pfg}, which corresponds to the upper
bound of $|m_{\beta \beta }|\leq (61-165)$ meV at 90\% C.L. For information about those other experiments see Refs.~\cite{Aalseth:2017btx,Alduino:2017ehq,Albert:2017owj,Arnold:2016bed}.
%Aalseth:2017bt,Alduino:2017ehq,Albert:2017owj,Arnold:2016bed}.

%%%%%%%%%%%%%%%%%%%%%%%%%

\section{Phenomenology}

\label{sec:pheno} %%%%%%%%%%%%%%%%%%%%%%%%%
\noindent As previously stated, the physical sterile
neutrino spectrum contains six almost degenerate TeV scale neutrinos, which mix the active
ones, with mixing angles of the order of $\frac{\left( {\bf M}_{1}\right) _{ij}}{%
\sqrt{2}y^{\left( N\right) }v_{\rho_1 }}$ \ ($i,j=1,2,3$). In return, these
new couplings induce one-loop level phenomena through which we may
impose constraints to our model.%\newline

In this section we discuss the implications of our model in the lepton
flavor violating decays and in the anomalous magnetic dipole moments of the
muon and electron.

%%%%%%%%%%%%%%%%%%%%%%%%%

\subsection{Charged LFV decays}

\label{sec:LFV} %%%%%%%%%%%%%%%%%%%%%%%%%
\noindent The heavy sterile neutrinos together with the W gauge bosons induce the one-loop level decay $l_{i}\rightarrow l_{j}\gamma$, whose corresponding 
branching ratio reads \cite%
{Ilakovac:1994kj,Deppisch:2004fa,Lindner:2016bgg}%
\begin{eqnarray}
&\mathcal{B}r\left( l_{i}\rightarrow l_{j}\gamma \right) =\frac{\alpha
_{W}^{3}s_{W}^{2}m_{l_{i}}^{5}}{256\pi ^{2}M_{W}^{4}\Gamma _{i}}\left\vert
G_{ij}\right\vert ^{2},
\end{eqnarray}
where $s_W = \sin (\theta_W)$,
\begin{eqnarray}
G_{ij} &=\sum_{k}\left( \mathbb{R}^{\ast }\right) _{ik}\left( \mathbb{R}%
\right) _{jk}G_{\gamma }\left( \frac{m_{N_{k}}^{2}}{M_{W}^{2}}\right) \simeq
2\left( {\bf R}_{1}{\bf R}_{1}^{T}\right) _{ij}G_{\gamma }\left( \frac{m_{N}^{2}}{%
M_{W}^{2}}\right) =\frac{\left( {\bf M}_{1}^{\ast }{\bf M}_{1}^{\dagger }\right) _{ij}}{%
m_{N}^{2}}G_{\gamma }\left( \frac{m_{N}^{2}}{M_{W}^{2}}\right) ,  \notag \\
&G_{\gamma } (z) = -\frac{2z^{3}+5z^{2}-z}{4\left( 1-z\right) ^{2}}-\frac{%
3z^{3}}{2\left( 1-z\right) ^{4}}\ln z,\hspace{0.7cm}\hspace{0.7cm}%
m_{N}=y^{\left( N\right) }v_{\rho_1},  \label{Brmutoegamma}
\end{eqnarray}%
and 
\begin{equation}
{\bf M}_{1}^{\ast }{\bf M}_{1}^{\dagger }=f^{2}\left( 
\begin{array}{ccc}
0 & x & 0 \\ 
y & 0 & rx \\ 
0 & ry & 0%
\end{array}%
\right) \left( 
\begin{array}{ccc}
0 & y & 0 \\ 
x & 0 & ry \\ 
0 & rx & 0%
\end{array}%
\right) =f^{2}\left( 
\begin{array}{ccc}
x^{2} & 0 & rxy \\ 
0 & r^{2}x^{2}+y^{2} & 0 \\ 
rxy & 0 & r^{2}y^{2}%
\end{array}%
\right) .
\end{equation}
Thus, the charged lepton flavor violating processes $\mu
\rightarrow e\gamma $, $\tau \rightarrow \mu \gamma $, and $\tau \rightarrow
e\gamma$ have the following branching ratios: 
\begin{eqnarray}
\mathcal{B}r\left( \mu \rightarrow e\gamma \right) &\simeq &0,\hspace{0.7cm}%
\hspace{0.7cm}\mathcal{B}r\left( \tau \rightarrow \mu \gamma \right) \simeq
0,  \notag \\
\mathcal{B}r\left( \tau \rightarrow e\gamma \right) &=&\frac{\alpha
_{W}^{3}s_{W}^{2}m_{\tau }^{5}r^{2}x^{2}y^{2}f^{4}}{256\pi
^{2}M_{W}^{4}\Gamma _{\tau }m_{N}^{4}}\left\vert G_{\gamma }\left( \frac{%
m_{N}^{2}}{M_{W}^{2}}\right) \right\vert ^{2},
\end{eqnarray}%
where $\Gamma _{\tau }=2.27\times 10^{-12}$ GeV is the tau decay width. On the
other hand, the upper experimental bound of the charged lepton flavor
violating process $\tau \rightarrow e\gamma $ is given by
\begin{equation}
\mathcal{B}r\left( \tau \rightarrow e\gamma \right) _{\max }^{\exp
}=3.3\times 10^{-8}.
\end{equation}

\begin{figure}[tbp]
\includegraphics[width=0.7\textwidth]{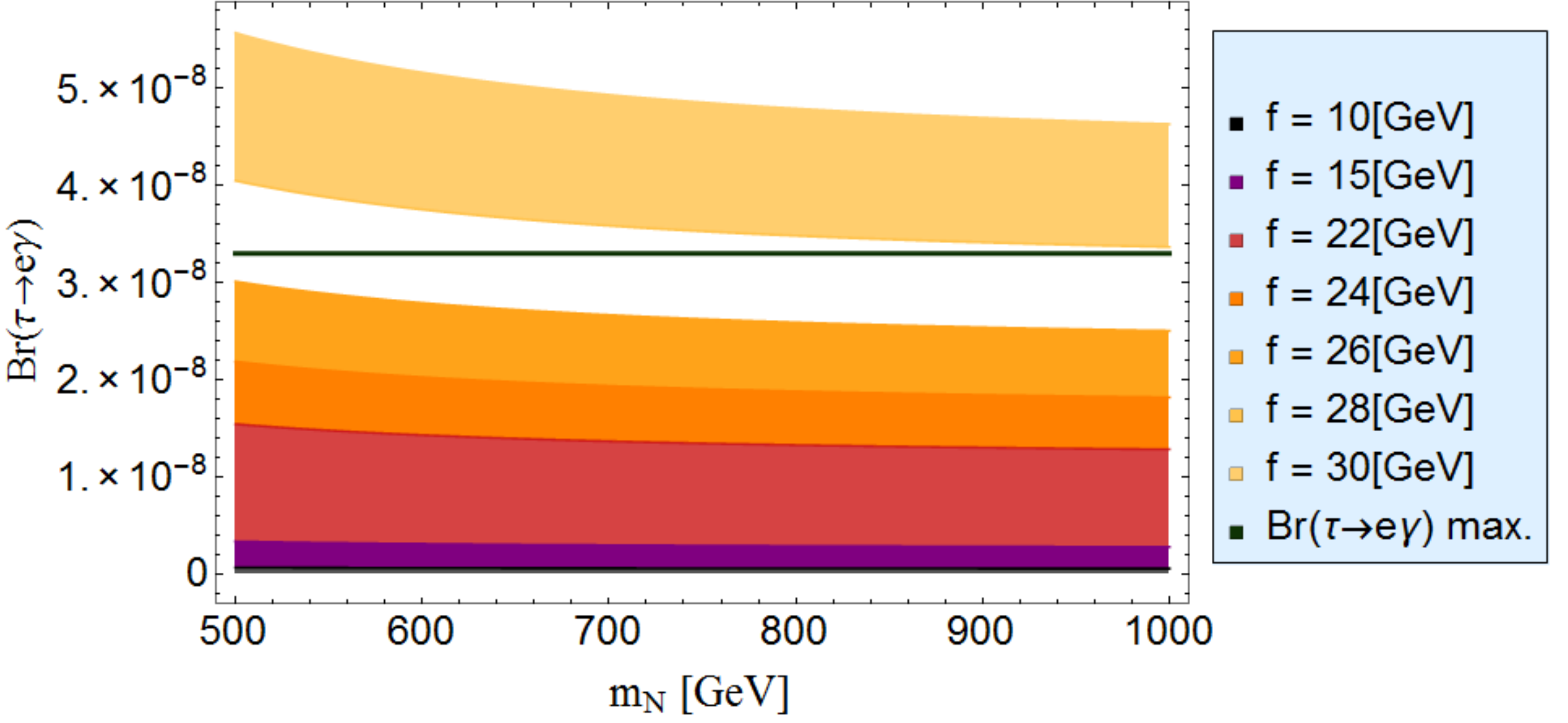}
\caption{Maximal and minimal branching ratios for the $\tau\to e\gamma$ decay as a function of the sterile neutrino mass $m_N$ for different values of $f$. The brown horizontal line corresponds to the upper bound $3.3\times 10^{-8}$ for the $\tau\to e\gamma$ branching ratio.}
\label{LFVplot}
\end{figure}

In Fig.~\ref{LFVplot} we display the maximal and minimal branching ratios for the $\tau\to e\gamma$ decay as functions of the sterile neutrino mass $m_N$ for different values of $f$. The sterile neutrino masses have been taken to range from $500$ GeV up to $1$ TeV. The brown horizontal line corresponds to the upper bound $3.3\times 10^{-8}$ for the $\tau\to e\gamma$ branching ratio. As seen from Fig.~\ref{LFVplot}, the obtained values for the branching ratio of $\tau \rightarrow e\gamma $ decay are below its experimental upper limit, for $f\lesssim 28$ GeV. Consequently, our model is compatible with the charged lepton flavor violating decay constraints
provided that $f\lesssim 28$ GeV.
%Besides that, it is worth mentioning that in a similar region of parameter space, one can obtain branching ratios for the $\tau\to e\gamma$ decay consistent with the projected sensitivity $3.3\times 10^{-9}$ of the forthcoming experiments.

\subsection{Contributions to $(g-2)_\protect\mu$}

\label{sec:Gminus2} %%%%%%%%%%%%%%%%%%%%%%%%%

\noindent The current discrepancy between the experimental and predicted
value is still inconclusive and amounts to $3.5$ standard deviations~\cite%
{Tanabashi:2018oca}, 
\begin{align}
\Delta a_\mu \equiv a_\mu^\text{exp} - a_\mu^\text{SM} = 268(63)(43)\times
10^{-11} \;,
\end{align}
where the errors at $1\,\sigma$ are from experiment and theory,
respectively. In the following, we consider the average value between the
theoretical and experimental error.

\begin{figure*}[tbp]
\includegraphics[scale=0.37]{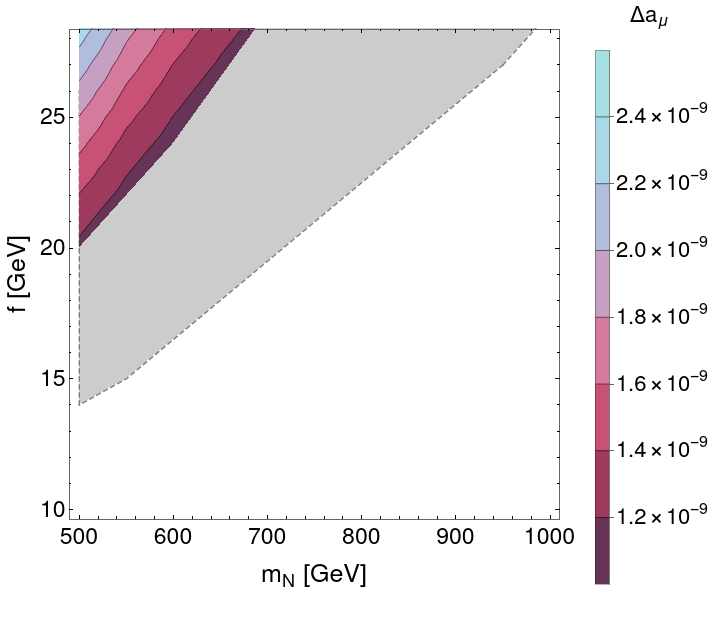}
\caption{Available parameter space in the $m_{N}-f$ plane accommodating $%
\Delta a_\protect\mu$ at 3$\,\protect\sigma$. The gray background considers
variations of the set of parameters in the range $0.3 \lesssim \{x,y\}
\lesssim 1$ and $-4.5 \lesssim r \lesssim 2.9$ while the colored bands
depict a particular benchmark scenario with $\{x,y\}=0.8$ and $r=-2.5$ where one may more easily
appreciate the dependence of $f$ and $m_N$ in $\Delta a_\protect\mu$.}
\label{fig:gminus2}
\end{figure*}

Contributions to $\Delta a_\mu$ arising from scenarios like this one where
the active neutrinos mix with heavy-right handed neutrinos have been already
computed. The relevant expression is given by~\cite%
{Leveille:1977rc,Lindner:2016bgg}, 
\begin{align}
\Delta a_\mu = \frac{-1}{8\pi^2} \kappa_\mu^2 \int^1_0 dz \sum_f \frac{|%
\mathcal{R}_{f\mu}^v|^2 P_3^+(z)+|\mathcal{R}_{f\mu}^a|^2 P_3^-(z)}{%
\epsilon_f^2 \kappa_\mu^2 (1-z) (1-\epsilon_f^{-2}z)+z} \;,
\end{align}
with 
\begin{align}
P_3^\pm (z) = -2z^2 (1+z\mp 2\epsilon_f) +\kappa_\mu^2 z
(1-z)(1\mp\epsilon_f)^2(z\pm\epsilon_f) \;,
\end{align}
and $\epsilon_f = \tfrac{m_{Nf}}{m_\mu}$ and $\kappa_\mu = \tfrac{m_\mu}{M_W}
$. In our particular case, the vector and axial-vector couplings to the $W$
bosons are identical and thus the expression is reduced to 
\begin{align} \label{eq:delta}
\Delta a_\mu = \frac{-1}{4\pi^2} \kappa_\mu^2 \sum_f |\mathcal{R}%
_{f\mu}^v|^2 \int_0^1 dz \frac{-z^2(1+z)+\kappa_\mu^2
z(1-z)[z+\epsilon_f^2(z-2)]} {\epsilon_f^2 \kappa_\mu^2 (1-z)
(1-\epsilon_f^{-2}z)+z} \;.
\end{align}
Notice that only two couplings contribute, $\mathcal{R}_{1\mu}^v$ and $%
\mathcal{R}_{3\mu}^v$. Both of them can be approximated to $\mathcal{R}%
_{1(3)\mu}^v \simeq \tfrac{f}{\sqrt{2}m_N}$ times order 1 parameters, $%
\{x,y,r\} \sim \mathcal{O}(1)$. Fig.~\ref{fig:gminus2} exemplifies the
available parameter space in the $m_{N}-f$ plane which accommodates $\Delta
a_\mu$ at 3$\,\sigma$. 
The gray background was obtained through variations of the set of parameters in the ranges $0.3 \lesssim \{x,y\}\lesssim 1$ and $-4.5\lesssim r \lesssim 2.9$ whereas the colored bands via a particular scenario where all order 1 parameters were fixed to $\{x, y\} = 0.8$ and $r = -2.5$.
Notice that the aforementioned range of parameters has been considered in order to show that our model can successfully accommodate the muon anomalous magnetic moment with order 1 dimensionless parameters and that such an observable can be used to set constraints on the dimensionful parameters $f$ and $m_N$. 

\subsection{Contributions to $(g-2)_e$}
\label{sec:Gminus2E} %%%%%%%%%%%%%%%%%%%%%%%%%

Recently, a new discrepancy between the experimental and predicted value in
the magnetic moment of the electron was found~\cite{Parker:2018vye},
\begin{align}
\Delta a_e \equiv a_e^\text{exp} - a_e^\text{SM} = - 88(36)\times 10^{-12} \;,
\end{align}
which amounts to $2.5\sigma$ standard deviations from the SM value. 

Contributions to $\Delta a_e$ may be similarly computed as in the previous
case. It is only necessary to replace the muon mass for the electron one in
 Eq.~\eqref{eq:delta}. In this case, variations of the set of parameters in the ranges $10 \text{ GeV} \lesssim f \lesssim 28 \text{ GeV}$, $0.5 \text{ TeV} \leq m_N \leq 1.0 \text{ TeV}$, $0.3
\lesssim \{x,y\} \lesssim 1$ and $-4.5 \lesssim r \lesssim 2.9$ lead to a full
available parameter space in the $m_{N}-f$ plane accommodating $\Delta
a_e$ at 3$\,\sigma$. However,
our model is only consistent with positive values for $\Delta a_e$ and predicts
it to be of the order of $10^{-16,\cdots,-13}$. Note that both anomalies can be 
simultaneously reproduced in the same range of parameters with order 1 dimensionless couplings for $14 \text{ GeV} \lesssim f \lesssim 28 \text{ GeV}$ and sterile neutrino masses in the range $0.5 \text{ TeV} \leq m_N \leq 1.0 \text{ TeV}$.

%%%%%%%%%%%%%%%%%%%%%%%%%

\section{Conclusions}

\label{sec:conclusions} %%%%%%%%%%%%%%%%%%%%%%%%%
\noindent We have proposed a viable low-scale seesaw model based on the $%
A_{4}$ family symmetry and other auxiliary cyclic symmetries, where the SM
particle spectrum is enlarged by the inclusion of several charged vectorlike 
fermions, right-handed Majorana neutrinos and scalar singlets,
consistent with the low energy SM fermion flavor data. The masses for the SM
charged fermions lighter than the top quark emerge from a universal seesaw
mechanism mediated by charged vectorlike fermions, whereas the small light
active neutrino masses are generated from an Inverse Seesaw mechanism. The
smallness of the $\mu$ parameter of the inverse seesaw, generated after the
spontaneous breaking of the discrete symmetries of the model, is attributed to a
right-handed neutrino nonrenormalizable Yukawa term. The spontaneous breaking of these discrete symmetries takes place at large
energies and gives rise to the observed SM fermion mass spectrum indicated by Eqs.~\eqref{eq:MtEW}-\eqref{eq:massHier} and fermionic mixing
parameters. Because of the discrete symmmetries of the model, the resulting leptonic mixing matrix corresponds to the experimentally observed deviation of the tripermuting
scenario and the SM charged lepton masses are linked with the scale of electroweak symmetry breaking through their power dependence on the Wolfenstein parameter $\lambda =0.225$, with $\mathcal{O}%
(1)$ coefficients. Furthermore, our model provides defined correlations between the two smallest quark mixing angles and between the Jarlskog invariant with
each of them, which is highly consistent with their allowed $1\sigma$ experimental values, except for the correlation between the smallest quark mixing angle and the Jarlskog invariant, where few points are within their $1\sigma$ experimentally allowed range. However, despite this small issue, most of the points in such correlations are inside their $2\sigma$ experimentally allowed range. In addition, from those correlations, we have found that our model prefers small values of the Jarlskog invariant compared to the latest fit from the PDG~\cite{Tanabashi:2018oca}, thus making a more precise measurement of such an invariant crucial to assess its viability. We have studied the implications of our model in the lepton
flavor violating decays and in the anomalous magnetic dipole moments of the
muon and electron. We have found that the $\mu \rightarrow e\gamma $ and $\tau
\rightarrow \mu \gamma $ are strongly suppressed in our model, whereas the $%
\tau \rightarrow e\gamma $ decay can attain values of the order of $10^{-8}$, which is within the reach of the
current sensitivity of the forthcoming charged lepton flavor violation
experiments. Furthermore, the obtained values for the branching ratio for
the $\tau \rightarrow e\gamma $ are lower than its current experimental
bound for a Dirac neutrino mass parameter $f$ lower than about $28$ GeV. Finally, we have
found that our model successfully accommodates the experimental values of the
anomalous magnetic dipole moments of the muon and electron for $14 \text{ GeV} \lesssim f \lesssim 28 \text{ GeV}$ and sterile neutrinos lighter than about $1$ TeV.  
In regards to the electron magnetic dipole moment, we have found that our model is only consistent with positive values for $\Delta a_e$ and predicts it to be of the order of $10^{-16,\cdots,-13}$. This implies that a more precise measurement of the electron magnetic dipole moment is crucial to confirm or rule out the model under consideration.

%%%%%%%%%%%%%%%%%%%%%%%%%

\section*{ACKNOWLEDGMENTS}

%%%%%%%%%%%%%%%%%%%%%%%%%
\noindent A.E.C.H. has received funding from Fondecyt (Chile), Grants
No.~1170803, CONICYT PIA/Basal FB0821. U.J.S.S. acknowledges financial support
in the early stage of the work from a DAAD One-Year Research Grant and
during the late stages from CONACYT-M\'{e}xico. U.J.S.S. is grateful to FCFM (BUAP) for hospitality during the completion of this work. A.E.C.H is very
grateful to Professor Hoang Ngoc Long for the warm hospitality at the Institute of Physics, Vietnam Academy of Science and Technology, where this work was finished. The authors are grateful to Rupert Coy for many useful commentaries on our paper and to Pablo Roig for suggesting us the study of the anomalous magnetic moment of the electron.

\appendix
%\vspace{-0.2cm}
\section{THE PRODUCT RULES FOR $A_4$}

\label{app:A4} %%%%%%%%%%%%%%%%%%%%%%%%%%%%
\noindent Alternating symmetry groups, $A_n$, describe the even permutations
of a given number, $n$, of indistinguishable objects. The smallest one is $%
A_{4}$. It has one triplet $\mathbf{3}$\ and three distinct one-dimensional $%
\mathbf{1}$, $\mathbf{1}^{\prime }$, and $\mathbf{1}^{\prime \prime }$
irreducible representations, satisfying the following product rules, 
\begin{eqnarray}
&&\hspace{18mm}\mathbf{3}\otimes \mathbf{3}=\mathbf{3}_{s}\oplus \mathbf{3}%
_{a}\oplus \mathbf{1}\oplus \mathbf{1}^{\prime }\oplus \mathbf{1}^{\prime
\prime },  \label{A4-singlet-multiplication} \\[0.12in]
&&\mathbf{1}\otimes \mathbf{1}=\mathbf{1},\hspace{5mm}\mathbf{1}^{\prime
}\otimes \mathbf{1}^{\prime \prime }=\mathbf{1},\hspace{5mm}\mathbf{1}%
^{\prime }\otimes \mathbf{1}^{\prime }=\mathbf{1}^{\prime \prime },\hspace{%
5mm}\mathbf{1}^{\prime \prime }\otimes \mathbf{1}^{\prime \prime }=\mathbf{1}%
^{\prime }.  \notag
\end{eqnarray}

Considering $\left( a_{1},a_{2},a_{3}\right) $ and $\left(
b_{1},b_{2},b_{3}\right) $ as basis vectors for two $A_{4}$-triplets $%
\mathbf{3}$, the following relations are fulfilled, 
\begin{eqnarray}
&&\left( \mathbf{3}\otimes \mathbf{3}\right) _{\mathbf{1}%
}=a_{1}b_{1}+a_{2}b_{2}+a_{3}b_{3},  \label{triplet-vectors} \\
&&\left( \mathbf{3}\otimes \mathbf{3}\right) _{\mathbf{3}_{s}}=\left(
a_{2}b_{3}+a_{3}b_{2},a_{3}b_{1}+a_{1}b_{3},a_{1}b_{2}+a_{2}b_{1}\right) ,\
\ \ \ \left( \mathbf{3}\otimes \mathbf{3}\right) _{\mathbf{1}^{\prime
}}=a_{1}b_{1}+\omega a_{2}b_{2}+\omega ^{2}a_{3}b_{3},  \notag \\
&&\left( \mathbf{3}\otimes \mathbf{3}\right) _{\mathbf{3}_{a}}=\left(
a_{2}b_{3}-a_{3}b_{2},a_{3}b_{1}-a_{1}b_{3},a_{1}b_{2}-a_{2}b_{1}\right) ,\
\ \ \left( \mathbf{3}\otimes \mathbf{3}\right) _{\mathbf{1}^{\prime \prime
}}=a_{1}b_{1}+\omega ^{2}a_{2}b_{2}+\omega a_{3}b_{3},  \notag
\end{eqnarray}%
where $\omega =e^{i\frac{2\pi }{3}}$. The representation $\mathbf{1}$ is
trivial, while the nontrivial $\mathbf{1}^{\prime }$ and $\mathbf{1}%
^{\prime \prime }$ are complex conjugate to each other.

\end{document}